# Proving Craig and Lyndon Interpolation Using Labelled Sequent Calculi *

Roman Kuznets

TU Wien

roman@logic.at

## Abstract

We have recently presented a general method of proving the fundamental logical properties of Craig and Lyndon Interpolation (IPs) by induction on derivations in a wide class of internal sequent calculi, including sequents, hypersequents, and nested sequents. Here we adapt the method to a more general external formalism of labelled sequents and provide sufficient criteria on the Kripke-frame characterization of a logic that guarantee the IPs. In particular, we show that classes of frames definable by quantifier-free Horn formulas correspond to logics with the IPs. These criteria capture the modal cube and the infinite family of transitive Geach logics.

## 1. Introduction

The Craig Interpolation Property (CIP) is one of the fundamental properties desired of a logic, with monographs [7] and conferences [5, 13] devoted to its study. It states that, for any valid fact $A \to B$ of the logic, there must exists an *inerpolant* $C$ in the common language of $A$ and $B$ such that both $A \to C$ and $C \to B$ are valid. The Lyndon Interpolation Property (LIP) strengthens the CIP by requiring that not only propositional atoms in $C$ but even their polarities be common to $A$ and $B$. The interpolation properties have numerous, well-established connections to both mathematics (e.g., to algebra via amalgamation) and computer science [16, 18]. In this paper, we consider modal logics based on classical propositional logic and understand *common language* to mean common propositional atoms.

One of the standard methods of proving both CIP and LIP, or IPs for short, is constructing an interpolant by induction on a derivation of (a representation of) $A \to B$ in a suitable analytic sequent calculus. Apart from its constructiveness, the method is also modular: if the sequent system is strengthened by an extra rule, only this additional rule needs to be checked to extend the IPs to the resulting stronger logic.

Until recently, a major weakness of the method was the limited expressivity of analytic sequent calculi. Recent advances extended the reach of the method to nested sequents (Fitting and Kuznets [6]) and hypersequents (Kuznets [11]). These results were unified and generalized to a wide range of internal sequent-like formalisms in (Kuznets [10]). In this paper, we adapt the method to the external formalism of labelled sequents, which is strictly more expressive than these internal calculi and was just recently shown in (Dyckhoff and Negri [4]) to be even more expressive than previously believed.

A great advantage of labelled sequents over hypersequents and nested sequents is the existence of general methods of generating sequent rules from first-order frame conditions for Kripke-complete logics. In this paper, we harness this strength by outlining sufficient criteria on the frame conditions to guarantee the CIP and LIP. Moreover, we describe an algorithm for constructing an interpolant of a formula $A \to B$ from a given derivation of $\mathsf{w} : A \Rightarrow \mathsf{w} : B$ in the labelled calculus for the logic.

Since the LIP implies the CIP, we will often write LIP to mean both CIP and LIP. The paper is structured as follows. In Sect. 2, we describe the formalism of labelled sequents and adapt the method of proving the LIP from internal sequent-like calculi to labelled sequents. In Sect. 3, we show how to construct an interpolant for all the labelled rules of the basic modal logic $\mathsf{K}$. In Sect. 4, we prove that all logics complete w.r.t. quantifier-free Horn formulas enjoy the LIP and argue that the restriction to Horn clauses is essential. In Sect. 5, we extend the results of Sect. 4 to labelled sequents with equality atoms. Sect. 7 recounts the most relevant advances in proof-theoretic methods of proving interpolation. In Sect. 6, we extend the method to Horn-like geometric rules. and apply our findings to the infinite family of Geach logics. Finally, in Sect. 8, we summarize our results and discuss future research.

## 2. Interpolation for Labelled Sequent Calculi

### 2.1 Labelled Sequent Calculi

Negri and von Plato [14, Sect. 11] describe how to translate frame conditions for modal logics into rules of labelled calculi.

**Definition 1** (Labelled sequent)**.** A *labelled sequent* is a figure $\Gamma \Rightarrow \Delta$ with $\Gamma$ and $\Delta$ being multisets[1] consisting of *labelled formulas* $\mathsf{w} : A$ and *relational atoms* $\mathsf{wRo}$, where $A$ is a modal formula in negation normal form (NNF)[2] and $\mathsf{w}$ and $\mathsf{o}$ are labels from a fixed countable set $\mathsf{W}$ of labels.

**Definition 2** (Kripke model)**.** A *Kripke frame* is a pair $(W, R)$ with a set $W \neq \varnothing$ and $R \subseteq W \times W$. A *Kripke model* is a triple $\mathcal{M} = (W, R, V)$ where $(W, R)$ is a Kripke frame and $V \colon \mathsf{Prop} \to 2^W$ is a function on the set $\mathsf{Prop}$ of propositional atoms. The satisfaction relation between worlds $w \in W$ and modal formulas is defined recursively: $\mathcal{M}, w \Vdash P$ iff $w \in V(P)$; $\mathcal{M}, w \Vdash \overline{P}$ iff $w \notin V(P)$; $\mathcal{M}, w \Vdash A \wedge B$ iff $\mathcal{M}, w \Vdash A$ and $\mathcal{M}, w \Vdash B$; $\mathcal{M}, w \Vdash A \vee B$ iff

* This material is based upon work supported by the Austrian Science Fund (FWF) Lise Meitner Grant M 1770-N25.

[1] We use multisets following [14], but the method of proving interpolation can also handle sequences and sets without any changes.

[2] Negation is restricted to propositional atoms. Primary connectives are $\wedge$, $\vee$, $\Box$, and $\Diamond$. Negation $\overline{A}$ is a function of a formula $A$ defined via De Morgan laws. $A \to B := \overline{A} \vee B$.

$$\mathsf{w}\!:\!P, \Gamma \Rightarrow \Delta, \mathsf{w}\!:\!P \qquad \mathsf{w}\!:\!\overline{P}, \Gamma \Rightarrow \Delta, \mathsf{w}\!:\!\overline{P}$$

$$\mathsf{w}\!:\!P, \mathsf{w}\!:\!\overline{P}, \Gamma \Rightarrow \Delta \qquad \Gamma \Rightarrow \Delta, \mathsf{w}\!:\!P, \mathsf{w}\!:\!\overline{P}$$

$$\mathsf{w}\!:\!\bot, \Gamma \Rightarrow \Delta \qquad \Gamma \Rightarrow \Delta, \mathsf{w}\!:\!\top$$

**Table 1.** Initial sequents

$$\dfrac{\mathsf{w}\!:\!A, \mathsf{w}\!:\!B, \Gamma \Rightarrow \Delta}{\mathsf{w}\!:\!A \wedge B, \Gamma \Rightarrow \Delta}\,L\wedge \qquad \dfrac{\Gamma \Rightarrow \Delta, \mathsf{w}\!:\!A \quad \Gamma \Rightarrow \Delta, \mathsf{w}\!:\!B}{\Gamma \Rightarrow \Delta, \mathsf{w}\!:\!A \wedge B}\,R\wedge$$

$$\dfrac{\mathsf{w}\!:\!A, \Gamma \Rightarrow \Delta \quad \mathsf{w}\!:\!B, \Gamma \Rightarrow \Delta}{\mathsf{w}\!:\!A \vee B, \Gamma \Rightarrow \Delta}\,L\vee \qquad \dfrac{\Gamma \Rightarrow \Delta, \mathsf{w}\!:\!A, \mathsf{w}\!:\!B}{\Gamma \Rightarrow \Delta, \mathsf{w}\!:\!A \vee B}\,R\vee$$

**Table 2.** Propositional rules for NNF

$$\dfrac{\mathsf{o}\!:\!A, \mathsf{w}\!:\!\Box A, \mathsf{wRo}, \Gamma \Rightarrow \Delta}{\mathsf{w}\!:\!\Box A, \mathsf{wRo}, \Gamma \Rightarrow \Delta}\,L\Box \qquad \dfrac{\mathsf{wRo}, \Gamma \Rightarrow \Delta, \mathsf{o}\!:\!A}{\Gamma \Rightarrow \Delta, \mathsf{w}\!:\!\Box A}\,R\Box$$

$$\dfrac{\mathsf{wRo}, \mathsf{o}\!:\!A, \Gamma \Rightarrow \Delta}{\mathsf{w}\!:\!\Diamond A, \Gamma \Rightarrow \Delta}\,L\Diamond \qquad \dfrac{\mathsf{wRo}, \Gamma \Rightarrow \Delta, \mathsf{w}\!:\!\Diamond A, \mathsf{o}\!:\!A}{\mathsf{wRo}, \Gamma \Rightarrow \Delta, \mathsf{w}\!:\!\Diamond A}\,R\Diamond$$

**Table 3.** Modal rules (label $\mathsf{o}$ is an *eigenvariable* in $L\Diamond$ and $R\Box$)

$\mathcal{M}, w \Vdash A$ or $\mathcal{M}, w \Vdash B$; $\mathcal{M}, w \Vdash \Box A$ iff $\mathcal{M}, u \Vdash A$ whenever $wRu$; $\mathcal{M}, w \Vdash \Diamond A$ iff $\mathcal{M}, u \Vdash A$ for some $wRu$.

The satisfaction relation is extended to labelled sequents in [14, Def. 11.25] (under the name of validity):

**Definition 3** (Labelled semantics)**.** An *interpretation* is a function $\llbracket\cdot\rrbracket\colon \mathsf{W} \to W$ from labels to worlds in a given Kripke model $\mathcal{M} = (W, R, V)$. A labelled sequent $\Gamma \Rightarrow \Delta$ is *forced under* $\llbracket\cdot\rrbracket$, written $\mathcal{M} \vDash \llbracket\Gamma \Rightarrow \Delta\rrbracket$, if the following holds: if $\mathcal{M}, \llbracket\mathsf{w}\rrbracket \Vdash A$ for each $\mathsf{w}\!:\!A \in \Gamma$ and $\llbracket\mathsf{w}\rrbracket R \llbracket\mathsf{o}\rrbracket$ for each $\mathsf{wRo} \in \Gamma$, then $\mathcal{M}, \llbracket\mathsf{u}\rrbracket \Vdash B$ for some $\mathsf{u} : B \in \Delta$. A sequent $\Gamma \Rightarrow \Delta$ is *valid in a class* $\mathcal{C}_{\mathsf{L}}$ *of Kripke models*, written $\mathcal{C}_{\mathsf{L}} \vDash \Gamma \Rightarrow \Delta$, if $\mathcal{M} \vDash \llbracket\Gamma \Rightarrow \Delta\rrbracket$ for each model $\mathcal{M} \in \mathcal{C}_{\mathsf{L}}$ and each interpretation $\llbracket\cdot\rrbracket$ into $\mathcal{M}$.

Note that relational atoms in the consequent play no role in this satisfaction relation. Since interpretations not respecting relational atoms in $\Gamma$ satisfy $\Gamma \Rightarrow \Delta$ trivially, we can simplify the definition to match more closely the semantics for internal calculi from [10] .

**Definition 4** (Good map)**.** Let $\mathcal{M} = (W, R, V)$ be a Kripke model. A *good* $\mathcal{M}$*-map* on a labelled sequent $\Gamma \Rightarrow \Delta$ is an interpretation into $\mathcal{M}$ such that $\llbracket\mathsf{w}\rrbracket R \llbracket\mathsf{o}\rrbracket$ for each $\mathsf{wRo} \in \Gamma$.

*Remark* 5. There is a cosmetic difference between good maps for labelled sequents and for multisequents from [10]: the latter are defined on a finite set of components present in a given sequent, whereas the former are defined on the infinite set $\mathsf{W}$ of all labels. Thus, strictly speaking, a good map should have been a restriction of a given interpretation to the labels occurring in the sequent. It is, however, clear that this difference is immaterial .

*Remark* 6. Similar to Def. 3, the relational atoms from $\Delta$ are ignored by good maps. In fact, it is noted in [14, Sect. 11.3(a)] that restricting the use of relational atoms to antecedents does not affect the completeness of the calculus.

**Lemma 7.** *$\mathcal{C}_{\mathsf{L}} \vDash \Gamma \Rightarrow \Delta$ iff $\mathcal{M} \vDash \llbracket\Gamma \Rightarrow \Delta\rrbracket$ for each model $\mathcal{M} \in \mathcal{C}_{\mathsf{L}}$ and each good $\mathcal{M}$-map $\llbracket\cdot\rrbracket$ on $\Gamma \Rightarrow \Delta$.*

The rules of the labelled calculus **SK** for the basic modal logic $\mathsf{K}$ can be found in Tables 1–3. This calculus is obtained by trivial modifications of the calculus **G3K** from [14, Table 11.5]. Given that our language in NNF is almost a sublanguage of the full language, the completeness of our calculus almost follows from that of [14]. The only new objects we have is $\overline{P}$ for propositional atoms and $\top$, which correspond to $\neg P$ and $\neg\bot$ respectively. Instead of applying negation rules on $\neg P$ and $\neg\bot$ leading to initial sequents in the full calculus, we add three new types of initial sequents for $\overline{P}$ and one for $\top$. We also omit initial sequents of the type $\mathsf{wRo}, \Gamma \Rightarrow \Delta, \mathsf{wRo}$ because, as noted earlier, they do not affect the completeness of the proof system. As this simple translation shows, our calculus is sound and complete for the logic $\mathsf{K}$ because the full calculus is.

### 2.2 Componentwise Interpolation Property

The idea of proving the LIP using labelled (and other advanced) sequents is to replace the formula-level interpolation statement with a sequent-based one and reduce the LIP to it. In this section, we present this sequent-based Componentwise Interpolation Property (CWIP). Instead of presenting it in the general format from [10], we adapt the notions and notation to labelled sequents, which simplifies and streamlines things. For instance,

**Definition 8** (Multiformula)**.** Each labelled formula $\mathsf{w} : C$ is a *multiformula*. If $\mho_1$ and $\mho_2$ are multiformulas, then $\mho_1 \ovee \mho_2$ and $\mho_1 \owedge \mho_2$ are also *multiformulas*.

Let $\llbracket\cdot\rrbracket$ be an interpretation into a model $\mathcal{M}$. A labelled formula $\mathsf{w}\!:\!C$ is *forced* by this interpretation, written $\mathcal{M} \vDash \llbracket\mathsf{w}\!:\!C\rrbracket$, iff $\mathcal{M}, \llbracket\mathsf{w}\rrbracket \Vdash C$. $\mathcal{M} \vDash \llbracket\mho_1 \ovee \mho_2\rrbracket$ ($\mathcal{M} \vDash \llbracket\mho_1 \owedge \mho_2\rrbracket$) iff $\mathcal{M} \vDash \llbracket\mho_i\rrbracket$ for some (each) $i = 1, 2$.

Thus, the external conjunction $\owedge$ and disjunction $\ovee$ on multiformulas behave classically. To define the Componentwise Interpolation Property, we use abbreviations:

**Definition 9.** Let $\mathcal{M}$ be a Kripke model and $\llbracket\cdot\rrbracket$ be a good $\mathcal{M}$-map on a labelled sequent $\Gamma \Rightarrow \Delta$. We write $\mathcal{M} \vDash \llbracket\mathsf{Ant}(\Gamma)\rrbracket$ if $\mathcal{M}, \llbracket\mathsf{w}\rrbracket \Vdash A$ for each $\mathsf{w} : A \in \Gamma$. We write $\mathcal{M} \vDash \llbracket\mathsf{Cons}(\Delta)\rrbracket$ if $\mathcal{M}, \llbracket\mathsf{o}\rrbracket \Vdash B$ for some $\mathsf{o}\!:\!B \in \Delta$.

**Definition 10** (Componentwise Interpolation Property, CWIP)**.** A multiformula $\mho$ is a (*componentwise*) *interpolant of a labelled sequent* $\Gamma \Rightarrow \Delta$, written $\Gamma \overset{\mho}{\Rightarrow} \Delta$, if the following conditions hold:

1. each label $\mathsf{w}$ occurring in $\mho$ must occur either in $\Gamma$ or in a labelled formula from $\Delta$;
2. each positive propositional atom $P$ occurring in $\mho$ must occur both in $\Gamma$ and $\Delta$;
3. each negative propositional atom $\overline{P}$ occurring in $\mho$ must occur both in $\Gamma$ and $\Delta$;
4. for each model $\mathcal{M} \in \mathcal{C}_{\mathsf{L}}$ and each good $\mathcal{M}$-map $\llbracket\cdot\rrbracket$ on $\Gamma \Rightarrow \Delta$, both implications are true:

$$\mathcal{M} \vDash \llbracket\mathsf{Ant}(\Gamma)\rrbracket \quad \text{implies} \quad \mathcal{M} \vDash \llbracket\mho\rrbracket \ , \tag{1}$$

$$\mathcal{M} \vDash \llbracket\mho\rrbracket \quad \text{implies} \quad \mathcal{M} \vDash \llbracket\mathsf{Cons}(\Delta)\rrbracket \ . \tag{2}$$

A labelled calculus **SL** has the CWIP iff every **SL**-derivable labelled sequent has an interpolant.

### 2.3 Reduction of Lyndon Interpolation to the CWIP

The LIP is proved by reduction to the CWIP. The reduction relies on three requirements on the bilateral connections among the logic $\mathsf{L}$, its labelled calculus **SL**, and its class $\mathcal{C}_{\mathsf{L}}$ of Kripke models, outlined in [10]. In order for the CWIP to hold, one further requirement is necessary. We will now formulate these four requirements for the case of labelled sequents and show that they easily follow from [14]. In order to have a hope of fulfilling the CWIP, we impose

**Requirement III.** *If* $\mathbf{SL} \vdash \Gamma \Rightarrow \Delta$*, then for each model* $\mathcal{M} \in \mathcal{C}_{\mathsf{L}}$ *and for each good* $\mathcal{M}$*-map* $\llbracket\cdot\rrbracket$ *on* $\Gamma \Rightarrow \Delta$*, either* $\mathcal{M}, \llbracket\mathsf{w}\rrbracket \nVdash A$ *for some* $\mathsf{w}\!:\!A \in \Gamma$ *or* $\mathcal{M}, \llbracket\mathsf{o}\rrbracket \Vdash B$ *for some* $\mathsf{o}\!:\!B \in \Delta$.

**Lemma 11.** *Requirement III is fulfilled for all modal labelled calculi we construct according to [14].*

*Proof.* Follows from [14, Th. 11.27] and Lemma 7. □

**Requirement I.** $\mathsf{L} \vdash A \to B$ *implies* $\mathbf{SL} \vdash \mathsf{w}: A \Rightarrow \mathsf{w}: B$ *for all* $\mathsf{w}$.

**Requirement II.** $A \vDash_{\mathcal{C}_\mathsf{L}} B$ *implies* $\mathsf{L} \vdash A \to B$.

**Requirement IV.** *For each labelled sequent containing no relational atoms and a single label* $\mathsf{w}$*, each model* $\mathcal{M} \in \mathcal{C}_\mathsf{L}$*, and each world* $w \in \mathcal{M}$*, all interpretations* $[\![\cdot]\!]$ *with* $[\![\mathsf{w}]\!] = w$ *are good* $\mathcal{M}$*-maps on this labelled sequent.*

*Proof.* We start with Req. II. $A \vDash_{\mathcal{C}_\mathsf{L}} B$ means that $\mathcal{M}, w \Vdash A$ implies $\mathcal{M}, w \Vdash B$ for all $\mathcal{M} \in \mathcal{C}_\mathsf{L}$ and all worlds $w \in \mathcal{M}$. In other words, $\mathcal{M}, w \Vdash \overline{A} \vee B$ for all $\mathcal{M} \in \mathcal{C}_\mathsf{L}$ and $w \in \mathcal{M}$. Thus, $\mathsf{L} \vdash A \to B$ by the completeness of $\mathsf{L}$ w.r.t. $\mathcal{C}_\mathsf{L}$.

To prove Req. I, note that $\mathsf{L} \vdash A \to B$ implies $\mathcal{M}, w \Vdash \overline{A} \vee B$ for all $\mathcal{M} \in \mathcal{C}_\mathsf{L}$ and $w \in \mathcal{M}$ by the soundness of $\mathsf{L}$ w.r.t. $\mathcal{C}_\mathsf{L}$. Thus, $\mathcal{M}, w \Vdash A$ implies $\mathcal{M}, w \Vdash B$ for all $\mathcal{M} \in \mathcal{C}_\mathsf{L}$ and $w \in \mathcal{M}$. Thus, for any model $\mathcal{M}$, label $\mathsf{w}$, and interpretation $[\![\cdot]\!]$ into $\mathcal{M}$, we have $\mathcal{M}, [\![\mathsf{w}]\!] \Vdash A$ implies $\mathcal{M}, [\![\mathsf{w}]\!] \Vdash B$. i.e., $\mathcal{M} \vDash [\![\mathsf{w}: A \Rightarrow \mathsf{w}: B]\!]$. Since $\mathsf{w}: A \Rightarrow \mathsf{w}: B$ is, therefore, valid in $\mathcal{C}_\mathsf{L}$, it follows from [14, Cor. 11.29] that $\mathbf{SL} \vdash \mathsf{w}: A \Rightarrow \mathsf{w}: B$.

Finally, Req. IV is trivial in the absence of relational atoms (the only possible restriction on good maps for a singleton label $\mathsf{w}$ could come from $\mathsf{w}R\mathsf{w} \in \Gamma$). □

*Remark* 12. The restriction of Req. IV to sequents without relational atoms is new compared to [10].

**Lemma 13.** *Each multiformula* $\mho$ *with a single label* $\mathsf{w}$ *can be replaced with a labelled formula* $\mathsf{w}: C$ *with the same positive and negative propositional atoms as* $\mho$ *such that* $\mathcal{M} \vDash [\![\mho]\!]$ *iff* $\mathcal{M} \vDash [\![\mathsf{w}: C]\!]$ *for any model* $\mathcal{M}$ *and interpretation* $[\![\cdot]\!]$ *into* $\mathcal{M}$.

*Proof.* By induction on the construction of $\mho$. The case when $\mho$ is a labelled formula is trivial. If $\mho_1$ and $\mho_2$ can be replaced with $\mathsf{w}: C_1$ and $\mathsf{w}: C_2$ respectively, then $\mho_1 \varovee \mho_2$ and $\mho_1 \varowedge \mho_2$ can be replaced with $\mathsf{w}: (C_1 \vee C_2)$ and $\mathsf{w}: (C_1 \wedge C_2)$ respectively. □

**Theorem 14** (Reduction of LIP to CWIP). *Let a logic* $\mathsf{L}$*, a multisequent proof system* $\mathbf{SL}$*, and a class of Kripke models* $\mathcal{C}_\mathsf{L}$ *satisfy all Reqs. I–IV. If* $\mathbf{SL}$ *enjoys the CWIP, then* $\mathsf{L}$ *enjoys the CIP.*

*Proof.* Assume that $\mathbf{SL}$ satisfies the CWIP and that $\mathsf{L} \vdash A \to B$. By Req. I, $\mathbf{SL} \vdash \mathsf{w}: A \Rightarrow \mathsf{w}: B$ for any $\mathsf{w} \in \mathsf{W}$, and this sequent has a componentwise interpolant $\mho$ by the CWIP. By Lemma 13, $\mathsf{w}: A \overset{\mathsf{w}:C}{\Longrightarrow} \mathsf{w}: B$ for some formula $C$. Each positive (negative) propositional atom in $C$ must occur in both $A$ and $B$. For any model $\mathcal{M} \in \mathcal{C}_\mathsf{L}$ and world $w \in \mathcal{M}$, the interpretation defined by $[\![\mathsf{o}]\!] := w$ for all $\mathsf{o} \in \mathsf{W}$ is a good $\mathcal{M}$-map on $\mathsf{w}: A \Rightarrow \mathsf{w}: B$ by Req. IV. In particular, $\mathcal{M} \vDash [\![\mathsf{Ant}(\mathsf{w}: A)]\!]$ implies $\mathcal{M} \vDash [\![\mathsf{w}: C]\!]$, i.e., $\mathcal{M}, w \Vdash A$ implies $\mathcal{M}, w \Vdash C$. Given the arbitrariness of $\mathcal{M}$ and $w$, we conclude that $A \vDash_{\mathcal{C}_\mathsf{L}} C$. It now follows from Req. II that $\mathsf{L} \vdash A \to C$. The proof of $\mathsf{L} \vdash C \to B$ is analogous. Thus, $C$ is a Lyndon (and Craig) interpolant of $A \to B$. □

*Remark* 15. The reduction uses a derivation of $\mathsf{w}: A \Rightarrow \mathsf{w}: B$, which cannot have relational atoms in the consequent of any labelled sequent from the proof tree. Thus, from now on, we assume that relational (and later equality) atoms are never present in consequents of labelled sequents.

The following notation will be useful:

**Definition 16.** Given an interpretation $[\![\cdot]\!]$ into a Kripke model $\mathcal{M}$, a sequence of labels $\vec{\mathsf{o}}$ from $\mathsf{W}$, and a sequence of world $\vec{u}$ from $\mathcal{M}$ of the same length we define a new interpretation $[\![\cdot]\!]_{\vec{\mathsf{o}}}^{\vec{u}}$ as follows:

$$[\![\mathsf{w}]\!]_{\vec{\mathsf{o}}}^{\vec{u}} := \begin{cases} [\![\mathsf{w}]\!] & \text{if } \mathsf{w} \neq \mathsf{o}_i \text{ for any } i, \\ u_i & \text{if } \mathsf{w} = \mathsf{o}_i. \end{cases}$$

We omit the vector arrow for sequences of length 1.

## 3. Interpolation for the Rules of SK

As for internal multisequents, it is easy to find interpolants for all initial sequents from Table 1:

| | |
|---|---|
| $\mathsf{w}: P, \Gamma \overset{\mathsf{w}:P}{\Longrightarrow} \Delta, \mathsf{w}: P$ | $\mathsf{w}: \overline{P}, \Gamma \overset{\mathsf{w}:\overline{P}}{\Longrightarrow} \Delta, \mathsf{w}: \overline{P}$ |
| $\mathsf{w}: P, \mathsf{w}: \overline{P}, \Gamma \overset{\mathsf{w}:\bot}{\Longrightarrow} \Delta$ | $\Gamma \overset{\mathsf{w}:\top}{\Longrightarrow} \Delta, \mathsf{w}: P, \mathsf{w}: \overline{P}$ |
| $\mathsf{w}: \bot, \Gamma \overset{\mathsf{w}:\bot}{\Longrightarrow} \Delta$ | $\Gamma \overset{\mathsf{w}:\top}{\Longrightarrow} \Delta, \mathsf{w}: \top$ |

**Table 4.** Interpolating initial sequents

Similar to the case of sequent derivations for propositional classical logic, single-premise propositional rules $L\wedge$ and $R\vee$ do not require changing an interpolant. The same behavior is exhibited by the modal rules $L\Box$ and $R\Diamond$, as well as by a wide range of rules generated from mathematical axioms. Thus, we define sufficient criteria for single-premise rules to preserve interpolants.

**Definition 17** (Local rules). A single-premise labelled sequent rule

$$\frac{\Gamma_p \Rightarrow \Delta_p}{\Gamma_c \Rightarrow \Delta_c}\,\mathsf{r} \tag{3}$$

is called *local for a class* $\mathcal{C}_\mathsf{L}$ *of Kripke models* if

1. for any Kripke model $\mathcal{M} \in \mathcal{C}_\mathsf{L}$, any good $\mathcal{M}$-map on $\Gamma_c \Rightarrow \Delta_c$ is also a good $\mathcal{M}$-map on $\Gamma_p \Rightarrow \Delta_p$,
2. each label from $\Gamma_p$ or from (a labelled formula in) $\Delta_p$ must occur either in $\Gamma_c$ or in (a labelled formula from) $\Delta_c$,
3. the sets of positive (negative) propositional atoms in $\Gamma_p$ and in $\Delta_p$are a subset of those in $\Gamma_c$ and in $\Delta_p$ respectively,
4. $\mathcal{M} \vDash [\![\mathsf{Ant}(\Gamma_c)]\!]$ implies $\mathcal{M} \vDash [\![\mathsf{Ant}(\Gamma_p)]\!]$ for any $\mathcal{M} \in \mathcal{C}_\mathsf{L}$ and good $\mathcal{M}$-map on $\Gamma_c \Rightarrow \Delta_c$, and
5. $\mathcal{M} \vDash [\![\mathsf{Cons}(\Delta_p)]\!]$ implies $\mathcal{M} \vDash [\![\mathsf{Cons}(\Delta_c)]\!]$ for any $\mathcal{M} \in \mathcal{C}_\mathsf{L}$ and good $\mathcal{M}$-map on $\Gamma_c \Rightarrow \Delta_c$.

**Lemma 18.** *The rules* $L\wedge$ *and* $R\vee$ *from Table 2 and* $L\Box$ *and* $R\Diamond$ *from Table 3 are local for any class of Kripke models.*

**Lemma 19** (Interpolating local rules)**.** *If a rule* $\mathsf{r}$ *from* (3) *is local for a class of models* $\mathcal{C}_\mathsf{L}$ *and* $\Gamma_p \overset{\mho}{\Rightarrow} \Delta_p$*, then* $\Gamma_c \overset{\mho}{\Rightarrow} \Delta_c$.

*Proof.* The conditions on relevant labels and on common propositional atoms for the conclusion of the rule are inherited from the premise by the definition of local rules. Consider a Kripke model $\mathcal{M} \in \mathcal{C}_\mathsf{L}$ and a good $\mathcal{M}$-map $[\![\cdot]\!]$ on $\Gamma_c \Rightarrow \Delta_c$. Assume first that $\mathcal{M} \vDash [\![\mathsf{Ant}(\Gamma_c)]\!]$. Then $\mathcal{M} \vDash [\![\mathsf{Ant}(\Gamma_p)]\!]$ and $[\![\cdot]\!]$ is a good $\mathcal{M}$-map on $\Gamma_p \Rightarrow \Delta_p$. Hence, $\mathcal{M} \vDash [\![\mho]\!]$. Assume now that $\mathcal{M} \vDash [\![\mho]\!]$. Since $[\![\cdot]\!]$ is a good $\mathcal{M}$-map on $\Gamma_p \Rightarrow \Delta_p$, it follows that $\mathcal{M} \vDash [\![\mathsf{Cons}(\Delta_p)]\!]$, from which we conclude that $\mathcal{M} \vDash [\![\mathsf{Cons}(\Delta_c)]\!]$. □

A similar standard argument shows that the conjunction (disjunction) of interpolants of the premises of the rule $R\wedge$ ($L\vee$) is an interpolant for the conclusion of the rule (see Table 5).

To complete a proof of the CWIP for **SK**, it remains to deal with the rules $L\Diamond$ and $R\Box$. The appropriate transformations are adapted from the analogous rules for multisequents in [10].

**Lemma 20** (Interpolating basic eigenvariable modal rules)**.** *Interpolant transformations for* $L\Diamond$ *and* $R\Box$ *from Table 3 are presented in Table 6. To use these transformations, the interpolant of the premise needs to be in DNF (CNF) for the rule* $L\Diamond$ *(*$R\Box$*), which is always possible to achieve given the classical nature of* $\varovee$ *and* $\varowedge$*.*

$$\frac{\Gamma \xRightarrow{\mho_1} \Delta, \mathsf{w}\colon A \quad \Gamma \xRightarrow{\mho_2} \Delta, \mathsf{w}\colon B}{\Gamma \xRightarrow{\mho_1 \owedge \mho_2} \Delta, \mathsf{w}\colon A \wedge B}\; R\wedge \qquad\qquad \frac{\mathsf{w}\colon A, \Gamma \xRightarrow{\mho_1} \Delta \quad \mathsf{w}\colon B, \Gamma \xRightarrow{\mho_2} \Delta}{\mathsf{w}\colon A \vee B, \Gamma \xRightarrow{\mho_1 \ovee \mho_2} \Delta}\; L\vee$$

**Table 5.** Interpolating binary propositional rules

$$\frac{\mathsf{wRo}, \mathsf{o}\colon A, \Gamma \xRightarrow{\bigovee_{i=1}^{n}\left(\bigowedge_{j=1}^{m_i} \mathsf{w}_{ij}\colon D_{ij} \;\owedge\; \bigowedge_{k=1}^{l_i} \mathsf{o}\colon C_{ik}\right)} \Delta}{\mathsf{w}\colon \Diamond A, \Gamma \xRightarrow{\bigovee_{i=1}^{n}\left(\bigowedge_{j=1}^{m_i} \mathsf{w}_{ij}\colon D_{ij} \;\owedge\; \mathsf{w}\colon\left(\Diamond \bigwedge_{k=1}^{l_i} C_{ik}\right)\right)} \Delta}\; L\Diamond \qquad \frac{\mathsf{wRo}, \Gamma \xRightarrow{\bigowedge_{i=1}^{n}\left(\bigovee_{j=1}^{m_i} \mathsf{w}_{ij}\colon D_{ij} \;\ovee\; \bigovee_{k=1}^{l_i} \mathsf{o}\colon C_{ik}\right)} \Delta, \mathsf{o}\colon A}{\Gamma \xRightarrow{\bigowedge_{i=1}^{n}\left(\bigovee_{j=1}^{m_i} \mathsf{w}_{ij}\colon D_{ij} \;\ovee\; \mathsf{w}\colon\left(\Box \bigvee_{k=1}^{l_i} C_{ik}\right)\right)} \Delta, \mathsf{w}\colon \Box A}\; R\Box$$

**Table 6.** Interpolating basic eigenvariable modal rules. In each rule, $\mathsf{w} \neq \mathsf{o}$, $\mathsf{w}_{ij} \neq \mathsf{o}$ for any $i, j$, and $\mathsf{o}$ occurs in neither $\Gamma$ nor $\Delta$.

*Proof.* We prove the statement for the rule $L\Diamond$, leaving $R\Box$ as an exercise by analogy. The common propositional atoms condition is clearly preserved. The only label that disappears in the conclusion sequent, $\mathsf{o}$, is also removed from the interpolant. Let $\mathcal{M} = (W, R, V)$ be a Kripke model and $[\![\cdot]\!]$ be a good $\mathcal{M}$-map on $\mathsf{w}\colon \Diamond A, \Gamma \Rightarrow \Delta$.

Assume $\mathcal{M} \vDash [\![\mathsf{Ant}(\mathsf{w}\colon \Diamond A, \Gamma)]\!]$. Since $\mathcal{M}, [\![\mathsf{w}]\!] \Vdash \Diamond A$, there is $u \in W$ such that $[\![\mathsf{w}]\!] R u$ and $\mathcal{M}, u \Vdash A$. For $[\![\cdot]\!]^u_{\mathsf{o}}$, which is a good $\mathcal{M}$-map on $\mathsf{wRo}, \mathsf{o}\colon A, \Gamma \Rightarrow \Delta$ because $\mathsf{o}$ does not occur in $\Gamma$, we have $\mathcal{M} \vDash [\![\mathsf{Ant}(\mathsf{wRo}, \mathsf{o}\colon A, \Gamma)]\!]^u_{\mathsf{o}}$. Thus, for the interpolant of the premise of $L\Diamond$ from Table 6 and some $1 \leq i \leq n$,

$$\mathcal{M} \vDash \left[\!\!\left[ \bigowedge_{j=1}^{m_i} \mathsf{w}_{ij} : D_{ij} \;\owedge\; \bigowedge_{k=1}^{l_i} \mathsf{o} : C_{ik} \right]\!\!\right]^u_{\mathsf{o}}, \tag{4}$$

in particular, $\mathcal{M}, u \Vdash C_{ik}$ for all $k = 1, \dots, l_i$. Given that $[\![\mathsf{w}]\!] R u$, we see that $\mathcal{M}, [\![\mathsf{w}]\!] \Vdash \Diamond \bigwedge_{k=1}^{l_i} C_{ik}$.[3] It follows that

$$\mathcal{M} \vDash \left[\!\!\left[ \bigowedge_{j=1}^{m_i} \mathsf{w}_{ij} : D_{ij} \;\owedge\; \mathsf{w} : \left(\Diamond \bigwedge_{k=1}^{l_i} C_{ik}\right) \right]\!\!\right]^u_{\mathsf{o}}.$$

Further, given that neither $\mathsf{w}$ nor any of $\mathsf{w}_{ij}$ is $\mathsf{o}$,

$$\mathcal{M} \vDash \left[\!\!\left[ \bigovee_{i=1}^{n} \left( \bigowedge_{j=1}^{m_i} \mathsf{w}_{ij} : D_{ij} \;\owedge\; \mathsf{w} : \left(\Diamond \bigwedge_{k=1}^{l_i} C_{ik}\right) \right) \right]\!\!\right], \tag{5}$$

which completes the proof of (1) for the conclusion of $L\Diamond$.

Assume now that (5) holds. Then

$$\mathcal{M} \vDash \left[\!\!\left[ \bigowedge_{j=1}^{m_i} \mathsf{w}_{ij} : D_{ij} \;\owedge\; \mathsf{w} : \left(\Diamond \bigwedge_{k=1}^{l_i} C_{ik}\right) \right]\!\!\right]$$

holds for some $1 \leq i \leq n$. In particular, $\mathcal{M}, [\![\mathsf{w}]\!] \Vdash \Diamond \bigwedge_{k=1}^{l_i} C_{ik}$. Thus, there is $u \in W$ such that $[\![\mathsf{w}]\!] R u$ and $\mathcal{M}, u \Vdash C_{ik}$ for all $k = 1, \dots, l_i$. Therefore, (4) holds for the interpolant of the premise for a good $\mathcal{M}$-map $[\![\cdot]\!]^u_{\mathsf{o}}$ on $\mathsf{wRo}, \mathsf{o}\colon A, \Gamma \Rightarrow \Delta$. It follows that $\mathcal{M} \vDash [\![\mathsf{Cons}(\Delta)]\!]^u_{\mathsf{o}}$. Given that $\mathsf{o}$ does not occur in $\Delta$, this is equivalent to $\mathcal{M} \vDash [\![\mathsf{Cons}(\Delta)]\!]$. ☐

**Corollary 21.** $\mathsf{K}$ *enjoys the LIP.*

[3] It also holds for $l_i = 0$: the empty conjunction is $\top$ and $\mathcal{M}, [\![\mathsf{w}]\!] \Vdash \Diamond\top$.

## 4. Mathematical Rules

The strength of labelled sequents is their versatility. In particular, [14] outlines methods of transforming *mathematical*, *geometrical*, and *co-geometrical* frame properties into labelled rules. In light of the recent successes of Dyckhoff and Negri [4] in geometrizing frame conditions, we concentrate on the first two types of properties, discussing mathematical frame conditions in this section and geometrical implications in the next one. By [14, Prop. 6.8],

**Lemma 22.** *Any classical quantifier-free property of Kripke frames can be represented as*

$$P_1 \wedge \ldots \wedge P_m \to Q_1 \vee \ldots \vee Q_n\,, \tag{6}$$

*where each $P_i$ and $Q_j$ is a relational atom.*

Based on this lemma, quantifier-free frame properties are also called *universal*. A mathematical property (6) corresponds to a rule

$$\frac{P_1, \dots, P_m, Q_1, \Gamma \Rightarrow \Delta \quad \dots \quad P_1, \dots, P_m, Q_n, \Gamma \Rightarrow \Delta}{P_1, \dots, P_m, \Gamma \Rightarrow \Delta}. \tag{7}$$

In addition, if substitution instances of several $P_i$'s are the same formula $P$, then the so-called *closure condition* requires adding variants of (7) that contract the extra copies of $P$.

Common examples of mathematical frame conditions yielding single-premise rules can be found in Table 7; we call such mathematical conditions *single-conclusion*. An example of a mathematical condition generating a two-premise rule is *connectedness* $wRo \wedge wRr \to oRr \vee rRo$:

$$\frac{\mathsf{oRr}, \mathsf{wRo}, \mathsf{wRr}, \Gamma \Rightarrow \Delta \quad \mathsf{rRo}, \mathsf{wRo}, \mathsf{wRr}, \Gamma \Rightarrow \Delta}{\mathsf{wRo}, \mathsf{wRr}, \Gamma \Rightarrow \Delta}\; Conn \tag{8}$$

(it also generates $Eucl^*$ from Table 7 by the closure condition). These rules are taken from [14, Table 11.6], where one can find a detailed discussion of their construction, as well as the closure conditions for transitivity and Euclideanness. In particular, it is noted that the rule $Trans^*$ is admissible and can be omitted, which we chose not to do because this rule presents no problem for our interpolation method.

There is little hope of establishing interpolation properties for logics whose mathematical frame conditions generate multi-premise rules. The pivotal counterexample is the logic $\mathsf{S4.3}$, complete with respect to transitive, reflexive, and connected frames. It was shown in Maksimova [12] that this logic violates the Craig Interpolation Property. Thus, there is no way to propagate the interpolation proof through the rule (8), even in the presence of transitivity and reflexivity, which help in other situations. Thus, we concentrate on mathematical axioms with $n = 1$, i.e., on Horn clauses.

| Frame property | Rule |
| --- | --- |
| Reflexivity $wRw$ | $\dfrac{\mathsf{wRw}, \Gamma \Rightarrow \Delta}{\Gamma \Rightarrow \Delta}\ \mathit{Ref}$ |
| Transitivity | $\dfrac{\mathsf{wRr}, \mathsf{wRo}, \mathsf{oRr}, \Gamma \Rightarrow \Delta}{\mathsf{wRo}, \mathsf{oRr}, \Gamma \Rightarrow \Delta}\ \mathit{Trans}$ |
| $wRo \land oRr \to wRr$ | $\dfrac{\mathsf{wRw}, \mathsf{wRw}, \Gamma \Rightarrow \Delta}{\mathsf{wRw}, \Gamma \Rightarrow \Delta}\ \mathit{Trans}^*$ |
| Euclideanness | $\dfrac{\mathsf{oRr}, \mathsf{wRo}, \mathsf{wRr}, \Gamma \Rightarrow \Delta}{\mathsf{wRo}, \mathsf{wRr}, \Gamma \Rightarrow \Delta}\ \mathit{Eucl}$ |
| $wRo \land wRr \to oRr$ | $\dfrac{\mathsf{oRo}, \mathsf{wRo}, \Gamma \Rightarrow \Delta}{\mathsf{wRo}, \Gamma \Rightarrow \Delta}\ \mathit{Eucl}^*$ |
| Symmetry $wRo \to oRw$ | $\dfrac{\mathsf{oRw}, \mathsf{wRo}, \Gamma \Rightarrow \Delta}{\mathsf{wRo}, \Gamma \Rightarrow \Delta}\ \mathit{Sym}$ |

**Table 7.** Common mathematical frame conditions with single-premise rules ($*$ marks rules added by the closure condition)

**Theorem 23.** *Modal logics enjoy the LIP if they are complete w.r.t. any class $\mathcal{C}_{\mathsf{L}}$ of frames described by Horn clauses of the forms*

$$w_1 R u_1 \land \ldots \land w_m R u_m \to vRz \ \text{ and } \tag{9}$$

$$w_1 R u_1 \land \ldots \land w_m R u_m \to \bot \tag{10}$$

*Proof.* By [14, Th. 11.27, Cor. 11.29], such logics are described by labelled calculi obtained by extending **SK** with rules of type

$$\frac{\mathsf{vRz}, \mathsf{w}_1\mathsf{Ru}_1, \ldots, \mathsf{w}_m\mathsf{Ru}_m, \Gamma \Rightarrow \Delta}{\mathsf{w}_1\mathsf{Ru}_1, \ldots, \mathsf{w}_m\mathsf{Ru}_m, \Gamma \Rightarrow \Delta} \tag{11}$$

(with, possibly, several contracted instances of this rule due to the closure condition) for each frame property (9) and initial sequents $\mathsf{w}_1\mathsf{Ru}_1, \ldots, \mathsf{w}_m\mathsf{Ru}_m, \Gamma \Rightarrow \Delta$ for each frame property (10). It is obvious that there are no good $\mathcal{M}$-maps on these initial sequents within the class $\mathcal{C}_{\mathsf{L}}$, making $\mathsf{w}_1 : \bot$ a (vacuous) interpolant.

As for the rules of type (11), it is obvious that the definition of CWIP is insensitive to contractions or duplications of both labelled formulas and relational atoms (both types of rules, though not present in the labelled calculi discussed, are clearly local rules for any class of Kripke models and, thus, require no change in the interpolant). Moreover, by the *subterm property* [14, Sect. 11.5], it is possible to restrict the applications of such a rule to the cases where both $\mathsf{v}$ and $\mathsf{z}$ occur in the conclusion of the rule (as usual, occurrences in relational atoms from $\Delta$, if any, do not count).[4] Thus, it is sufficient to process all rules of type (11) with the subterm property, which we do by showing them to be local for $\mathcal{C}_{\mathsf{L}}$. Given that labelled formulas remain unchanged by (11), Clauses 3–5 of Def. 17 are trivially satisfied. To see that Clause 2 is satisfied, note that $\Delta_p = \Delta_c = \Delta$ and the only potentially new labels in $\Gamma_p$ are $\mathsf{v}$ and $\mathsf{z}$, which occur in the conclusion of the rule by our restriction based on the subterm property.

It remains to demonstrate Clause 1. Let $\mathcal{M} = (W, R, V) \in \mathcal{C}_{\mathsf{L}}$. Any good $\mathcal{M}$-map $[\![\cdot]\!]$ on the conclusion must satisfy all relational atoms from $\Gamma_c$, i.e., those from $\Gamma$, as well as satisfy $[\![\mathsf{w}_i]\!] R [\![\mathsf{u}_i]\!]$ for each $i = 1, \ldots, m$. Given that $\mathcal{M} \in \mathcal{C}_{\mathsf{L}}$, it follows from (9) that $[\![\mathsf{v}]\!] R [\![\mathsf{z}]\!]$ is satisfied.Thus, all relational atoms from $\Gamma_p$ are satisfied by $[\![\cdot]\!]$, making it a good $\mathcal{M}$-map on the premise of the rule.

Thus, any rule (11) generated from a Horn clause is local for $\mathcal{C}_{\mathsf{L}}$ (modulo the subterm property). The statement of the theorem now follows from Tables 4–5 and Lemmas 19–20. $\square$

[4] Note that requiring them to occur among $\mathsf{w}_i$'s and $\mathsf{u}_i$'s would be too strong as it would exclude, e.g., the reflexivity property.

**Corollary 24.** *Modal logics enjoy the LIP if they are complete w.r.t. any class of Kripke models defined by any combination of the following properties:*

- *reflexivity, transitivity, symmetry, Euclideanness, and* shift reflexivity *($wRo \to oRo$) from [8, Sect. 8];*
- *$(1, m)$-transitivity ($w_0 R w_1 \land \ldots \land w_{m-1} R w_m \to w_0 R w_m$) defined by analogy with $m$-transitivity, e.g., from [17];*
- *shift transitivity: $wRo \land oRu \land uRv \to oRv$;*
- *shift symmetry: $wRo \land oRu \to uRo$;*
- *shift Euclideanness: $wRo \land oRu \land oRv \to uRv$;*
- *any property obtained by replacing the "shift" condition $wRo$ above by an arbitrary conjunction of relational atoms;*
- *irreflexive discreteness: $wRv \to \bot$.*

*These logics include* $\mathsf{T}$, $\mathsf{KB}$, $\mathsf{K4}$, $\mathsf{K5}$, $\mathsf{S4}$, $\mathsf{Verum}$, $\mathsf{S5}$, $\mathsf{K4B}$, *and* $\mathsf{K4}_{1,m}$ *from the list of standard normal modal logics [3, Table 4.2].*

## 5. Labelled Sequents with Equality

As discussed in [14, Sect. 11.6], the formalism of labelled sequents can be enriched by equality atoms $\mathsf{w} = \mathsf{o}$ without affectingthe completeness results if the sequent calculi are appended by the rules in Table 8. Equality atoms can be treated by labelled calculi and by interpolation method the same way as relational atoms. In particular, the subterm property can be extended to systems with equality, equality atoms can be disallowed in the consequents, etc. If the definition of *good $\mathcal{M}$-maps* on $\Gamma \Rightarrow \Delta$ is appended by the condition that $[\![\mathsf{w}]\!] = [\![\mathsf{o}]\!]$ for each equality atom $\mathsf{w} = \mathsf{o}$ from $\Gamma$, it is easy to see that

**Lemma 25.** *All rules from Table 8 restricted by the subterm property are local for any class of Kripke models.*

Further, it is easy to see that the proof of Lemma 19 applies to labelled calculi with equality as is. Using the same construction of labelled rules from Horn clauses as (11) in the previous section, we can prove that all such rules with equality restricted by the subterm property are local for the respective class of frames in the same way. This yields

**Theorem 26.** *Modal logics enjoy the LIP if they are complete w.r.t. a class $\mathcal{C}_{\mathsf{L}}$ of frames described by Horn clauses of the form*

$$w_1 R u_1 \land \ldots \land w_m R u_m \quad \to \quad Q \ .$$

*where $Q$ can be $vRz$, or $v = z$, or $\bot$.*

**Corollary 27.** *Modal logics enjoy the LIP if they are complete w.r.t. any class of Kripke models defined by any combination of the properties described in Cor. 24 and of the following properties:*

- *functionality ($wRo \land wRu \to o = u$) from [8, Sect. 8];*
- *reflexive or irreflexive discreteness: $wRu \to w = u$.*

*In particular, the list of logics with the LIP proved using labelled sequents is extended by the logic or reflexive discrete frames* $\mathsf{Triv}$ *from [3, Table 4.2].*

## 6. Geometric Rules

Although the methods used in the preceding sections are adopted without much ado from the multisequent setting of [10], still the results obtained via labelled sequents are much stronger due to the expressivity of the latter and the method of generating local labelled rules from mathematical axioms.

In this section, we go beyond the interpolant transformations used in [10] in order to tackle labelled rules generated from frame conditions given by the so-called *geometric implications*, or rather by their canonical forms, according to [14, Table 8.1].

$$\frac{\mathsf{w}=\mathsf{w},\Gamma\Rightarrow\Delta}{\Gamma\Rightarrow\Delta}\,\textit{Eq-Ref}\qquad \frac{\mathsf{o}=\mathsf{r},\mathsf{w}=\mathsf{o},\mathsf{w}=\mathsf{r},\Gamma\Rightarrow\Delta}{\mathsf{w}=\mathsf{o},\mathsf{w}=\mathsf{r},\Gamma\Rightarrow\Delta}\,\textit{Eq-Trans}\qquad \frac{\mathsf{o}:A,\mathsf{w}=\mathsf{o},\mathsf{w}:A,\Gamma\Rightarrow\Delta}{\mathsf{w}=\mathsf{o},\mathsf{w}:A,\Gamma\Rightarrow\Delta}\,\textit{Repl}$$

$$\frac{\mathsf{oRr},\mathsf{w}=\mathsf{o},\mathsf{wRr},\Gamma\Rightarrow\Delta}{\mathsf{w}=\mathsf{o},\mathsf{wRr},\Gamma\Rightarrow\Delta}\,\textit{Repl}_{R_1}\qquad \frac{\mathsf{wRr},\mathsf{o}=\mathsf{r},\mathsf{wRo},\Gamma\Rightarrow\Delta}{\mathsf{o}=\mathsf{r},\mathsf{wRo},\Gamma\Rightarrow\Delta}\,\textit{Repl}_{R_2}$$

**Table 8.** Rules for equality atoms

**Definition 28** (Canonical form of geometric implication)**.** A *canonical geometric implication* has the form

$$P_1 \wedge \dots P_m \to \exists \vec{y_1} M_1 \vee \dots \vee \vec{y_n} M_n$$

where each $P_i$ is a relational atom, each $M_j$ is a conjunction of relational atoms, and each $\vec{y_j}$ is a sequence of variables over worlds that do not occur in any of $P_i$'s.

Given that our method must fail for $n > 1$ even in the quantifier-free case, we only consider the case of $n = 1$ (the case of $n = 0$ coincides with the mathematical axioms considered earlier). Thus, we look at frame conditions of the form

$$P_1 \wedge \dots \wedge P_m \to \exists \vec{y}\big(Q_1(\vec{y}) \wedge \dots \wedge Q_l(\vec{y})\big)\ , \qquad (12)$$

which correspond to the rules in Table 9. Since we are unable, at this point, to provide a general proof of the CIP for such rules, we start with a special case.

### 6.1 Telescopic Rules

**Definition 29** (Telescopic axioms)**.** We abbreviate $T(x,\vec{y}) := xRy_1 \wedge y_1Ry_2 \wedge \dots \wedge y_{k-1}Ry_k$ for $\vec{y} = \langle y_1,\dots,y_k\rangle$ and call it a *telescope*. A single-conclusion geometric axiom (12) is called *telescopic* if it only adds new variables in one or several disjoint telescopes, i.e., if it has the form

$$\bigwedge_{i=1}^{m} w_iRo_i \to \exists \vec{y}^1 \dots \exists \vec{y}^a \left(\bigwedge_{j=1}^{n} u_jRv_j \wedge \bigwedge_{l=1}^{a} T(x_l,\vec{y}^l)\right) \qquad (13)$$

where no member of $\vec{y}^l$ coincides with any of $w_i$, $o_i$, $u_j$, $v_j$, or $x_l$ and all members of $\vec{y}^l$'s are pairwise distinct.[5]

The simplest and most familiar example of a telescopic axiom is *seriality*: $\exists o wRo$. The corresponding rule is $\dfrac{\mathsf{wRo},\Gamma\Rightarrow\Delta}{\Gamma\Rightarrow\Delta}\,\textit{Ser}$ where $\mathsf{o} \neq \mathsf{w}$ and $\mathsf{o}$ does not occur in $\Gamma$ or $\Delta$. The same subterm property discussed above allows to restrict the use of this rule to instances with $\mathsf{w}$ occuring either in $\Gamma$ or in $\Delta$. More generally, the rules from Table 9 can be restricted to instances with all label variables other than $\vec{\mathsf{z}}$ occurring in one of $P_i$'s, in $\Gamma$, or in $\Delta$.

Note that a multitelescopic axiom (13) can be equivalently represented as a conjunction of one mathematical axiom and several monotelescopic axioms:

$$\left(\bigwedge_{i=1}^{m} w_iRo_i \to \bigwedge_{j=1}^{n} u_jRv_j\right) \wedge \bigwedge_{l=1}^{a}\left(\bigwedge_{i=1}^{m} w_iRo_i \to \exists \vec{y}^l T(x_l,\vec{y}^l)\right).$$

Thus, it is sufficient to show interpolation for rules

$$\frac{\mathsf{xRy}_1,\mathsf{y}_1\mathsf{Ry}_2,\dots,\mathsf{y}_{k-1}\mathsf{Ry}_k,\mathsf{w}_1\mathsf{Ro}_1,\dots,\mathsf{w}_m\mathsf{Ro}_m,\Gamma\Rightarrow\Delta}{\mathsf{w}_1\mathsf{Ro}_1,\dots,\mathsf{w}_m\mathsf{Ro}_m,\Gamma\Rightarrow\Delta} \qquad (14)$$

where $\mathsf{x}$ occurs in the conclusion and $\mathsf{y}_1,\dots,\mathsf{y}_k$ are pairwise distinct eigenvariables, i.e., for rules generated from axioms

$$\bigwedge_{i=1}^{m} w_iRo_i \to \exists y_1 \dots \exists y_k \big(xRy_1 \wedge y_1Ry_2 \wedge \dots \wedge y_{k-1}Ry_k\big). \qquad (15)$$

**Theorem 30.** *For a class of Kripke models $\mathcal{C}_{\mathsf{L}}$ satisfying* (15)*, an interpolant transformation for the rule* (14) *is shown in Table 10.*[6]

*Proof.* It is easy to see that the common propositional atoms condition is preserved. The only labels that disappears from the conclusion sequent are $\mathsf{y}_j$'s and they are also removed from the interpolant. Let $\mathcal{M} = (W,R,V)$ be any Kripke model from $\mathcal{C}_{\mathsf{L}}$ and $[\![\cdot]\!]$ be a good $\mathcal{M}$-map on $\mathsf{w}_1\mathsf{Ro}_1,\dots,\mathsf{w}_m\mathsf{Ro}_m,\Gamma\Rightarrow\Delta$.

Assume that $\mathcal{M} \vDash [\![\mathsf{Ant}(\mathsf{w}_1\mathsf{Ro}_1,\dots,\mathsf{w}_m\mathsf{Ro}_m,\Gamma)]\!]$. Since $[\![\mathsf{w}_c]\!]R[\![\mathsf{o}_c]\!]$ for $c = 1,\dots,m$, it follows from (15) that there are worlds $y_j \in W$ such that $[\![\mathsf{x}]\!]Ry_1$, …, $y_{k-1}Ry_k$. Since $\mathsf{y}_j$ don't occur in $\Gamma$ and are distinct from $\mathsf{w}_c$ and $\mathsf{o}_c$, $[\![\cdot]\!]^{\vec{y}}_{\vec{\mathsf{y}}}$ is a good $\mathcal{M}$-map on $\mathsf{xRy}_1,\dots,\mathsf{y}_{k-1}\mathsf{Ry}_k,\mathsf{w}_1\mathsf{Ro}_1,\dots,\mathsf{w}_m\mathsf{Ro}_m,\Gamma\Rightarrow\Delta$ and $\mathcal{M} \vDash [\![\mathsf{Ant}(\mathsf{xRy}_1,\mathsf{y}_1\mathsf{Ry}_2,\dots,\mathsf{y}_{k-1}\mathsf{Ry}_k,\mathsf{w}_1\mathsf{Ro}_1,\dots,\mathsf{w}_m\mathsf{Ro}_m,\Gamma)]\!]^{\vec{y}}_{\vec{\mathsf{y}}}$. Thus, for the premise interpolant from Table 10 and some $1 \le i \le n$,

$$\mathcal{M} \vDash \left[\!\!\left[\bigotimes_{b=1}^{m_i} \mathsf{u}_{ib} : D_{ib} \ \otimes\ \bigotimes_{j=1}^{k} \mathsf{y}_j : C_{ij}\right]\!\!\right]^{\vec{y}}_{\vec{\mathsf{y}}}\ , \qquad (16)$$

in particular, $\mathcal{M}, y_j \Vdash C_{ij}$ for all $j = 1,\dots,k$ for this $i$. It is easy to show by induction that

$$\mathcal{M}, y_j \Vdash C_{ij} \wedge \Diamond(C_{i,j+1} \wedge \Diamond(\dots \wedge \Diamond(C_{i,k-1} \wedge \Diamond C_{ik})\dots))$$

culminating in $\mathcal{M}, [\![\mathsf{x}]\!] \Vdash \mathbb{T}_i$ for

$$\mathbb{T}_i := \Diamond(C_{i,1} \wedge \Diamond(C_{i,2} \wedge \Diamond(\dots \wedge \Diamond(C_{i,k-1} \wedge \Diamond C_{ik})\dots)))$$

Given that neither of $\mathsf{u}_{ib}$ coincides with any of $\mathsf{y}_j$, it follows that

$$\mathcal{M} \vDash \left[\!\!\left[\bigotimes_{i=1}^{n}\left(\bigotimes_{b=1}^{m_i} \mathsf{u}_{ib} : D_{ib} \ \otimes\ \mathsf{x} : \mathbb{T}_i\right)\right]\!\!\right]\ . \qquad (17)$$

Assume now that (17) holds. Then

$$\mathcal{M} \vDash \left[\!\!\left[\bigotimes_{b=1}^{m_i} \mathsf{u}_{ib} : D_{ib} \ \otimes\ \mathsf{x} : \mathbb{T}_i\right]\!\!\right]$$

holds for some $1 \le i \le n$. In particular, $\mathcal{M}, [\![\mathsf{x}]\!] \Vdash \mathbb{T}_i$. Thus, there exist worlds $y_j \in W$ such that $[\![\mathsf{x}]\!]Ry_1$, $y_1Ry_2$, …, $y_{k-1}Ry_k$ and $\mathcal{M}, y_j \Vdash C_{ij}$ for all $j = 1,\dots,k$. Again, $[\![\cdot]\!]^{\vec{y}}_{\vec{\mathsf{y}}}$ is a good $\mathcal{M}$-map on $\mathsf{xRy}_1,\mathsf{y}_1\mathsf{Ry}_2,\dots,\mathsf{y}_{k-1}\mathsf{Ry}_k,\mathsf{w}_1\mathsf{Ro}_1,\dots,\mathsf{w}_m\mathsf{Ro}_m,\Gamma\Rightarrow\Delta$ for which (16) holds for the interpolant of the premise. It follows that $\mathcal{M} \vDash [\![\mathsf{Cons}(\Delta)]\!]^{\vec{y}}_{\vec{\mathsf{y}}}$. Given that none of $\mathsf{y}_j$ occurs in $\Delta$, this is equivalent to $\mathcal{M} \vDash [\![\mathsf{Cons}(\Delta)]\!]$. □

[5] Some of the $P_i$'s can, in principle, be equality rather than relational atoms. However, any equality atom $w_i = o_i$ is easy to remove by substituting $w_i$ for $o_i$ in the axiom.

[6] Note that here we also collect all labelled formulas with the same eigenvariable $\mathsf{y}_j$ into one labelled formula by transforming $\mathsf{v} : A \otimes \mathsf{v} : B$ into $\mathsf{v} : (A \wedge B)$ if more than one formula is labelled $\mathsf{y}_j$ or by adding $\mathsf{y}_j : \top$ if no formula is labelled $\mathsf{y}_j$ in a disjunct.

$$\frac{Q_1(\vec{\mathsf{z}}), \dots, Q_l(\vec{\mathsf{z}}), P_1, \dots, P_m, \Gamma \Rightarrow \Delta}{P_1, \dots, P_m, \Gamma \Rightarrow \Delta}$$

**Table 9.** The geometric rule scheme: *eigenvariables* $\vec{\mathsf{z}}$ do not occur in $P_i$'s, $\Gamma$, or $\Delta$.

**Example 31.** For any class consisting only of models with serial frames, assuming that $\mathsf{w}$ occurs in $\Gamma$ or $\Delta$, $\mathsf{o}$ does not, and $\mathsf{o}$ does not coincide with any of $\mathsf{u}_{ib}$,

$$\dfrac{\mathsf{wRo}, \Gamma \xRightarrow{\bigotimes_{i=1}^{n}\left(\bigotimes_{b=1}^{m_i} \mathsf{u}_{ib}{:}D_{ib} \otimes \mathsf{o}{:}C_i\right)} \Delta}{\Gamma \xRightarrow{\bigotimes_{i=1}^{n}\left(\bigotimes_{b=1}^{m_i} \mathsf{u}_{ib}{:}D_{ib} \otimes \mathsf{w}{:}\Diamond C_i\right)} \Delta}\ Ser\ ,$$

which is essentially the same transformation as used for $L\Diamond$.

**Corollary 32.** *Modal logics enjoy the LIP if they are complete w.r.t. any class of Kripke models defined by any combination of the properties described in Cors. 24 and 27 and of the following properties*

- *seriality;*
- *shift seriality:* $wRu \to \exists o uRo$*;*
- *any property obtained by replacing the "shift" condition* $wRu$ *above by an arbitrary conjunction of relational atoms.*

*In particular, the list of logics with the LIP proved using labelled sequents is extended by the logics* D *and* D4 *from [3, Table 4.2].*

### 6.2 Geometric Rules

While $\Diamond$ helps describe an accessible world, more complex configurations of eigenvariables are hard to describe by modal formulas. Consider the two remaining frame properties from [8, Sect. 8] that are not yet susceptible to our methods:

- *density*: $wRu \to \exists v(wRv \wedge vRu)$,
- *convergence*: $wRu \wedge wRo \to \exists v(uRv \wedge oRv)$,

both of which are single-conclusion geometric implications in canonical form. It is not clear which formulas are to be true at $w$, $r$, and $u$ in order to ensure that the interpolant information from the conclusion can be lifted to the premise. For instance, for the case of convergence, $\mathsf{u} : \Diamond C$ only describes a world satisfying $C$ and accessible from $[\![\mathsf{u}]\!]$. It is not clear how to pinpoint a world satisfying $C$ and simultaneously accessible from two given worlds $[\![\mathsf{u}]\!]$ and $[\![\mathsf{o}]\!]$. Indeed, $\mathsf{u} : \Diamond C \otimes \mathsf{o} : \Diamond C$ only implies that each of the two worlds has an accessible world, $u'$ and $o'$ respectively, satisfying $C$ but cannot guarantee that $u' = o'$. To overcome this difficulty, we use a convergence-like property to find a third $C$-world $v$ accessible from both $u'$ and $o'$ and a transitivity-like property to ensure that $v$ is directly accessible from both original worlds $[\![\mathsf{u}]\!]$ and $[\![\mathsf{o}]\!]$.

In this section, we outline general conditions and an interpolant transformation that allow to carry the interpolation proof through geometric rules with eigenvariables not forming disjoint telescopes. While the conditions themselves are a bit technical, they can be viewed as weakened forms of transitivity and convergence adapted to the particulars of a given sequent rule. In particular, both density and convergence become amenable to our method in presence of some additional frame properties.

Purely for simplicity of notation, we restrict the atoms $P_i$ and $Q_j(\vec{y})$ appearing in single-conclusion canonical geometric implications (12) to relational atoms. It is easy to see that equality atoms can be removed by substitutions applied to the frame condition, sometimes producing a quantifier-free condition equivalent to the given geometric one with equality. Moreover, we assume that every $Q_j(\vec{y})$ contains an occurrence of an eigenvariable from $\vec{y}$ because otherwise it can be outsourced to a separate quantifier-free condition, all of which we treated in the preceding sections. Thus, we plan to demonstrate interpolation for frame conditions of the form

$$\bigwedge_{i=1}^{m} w_i R o_i \to \exists y_1 \dots \exists y_k \bigwedge_{j=1}^{l} x_j R e_j\ , \tag{18}$$

where $\{x_j, e_j\} \cap \{y_1, \dots, y_k\} \neq \varnothing$ for each $1 \leq j \leq l$ and $\{y_1, \dots, y_k\} \subseteq \{x_1, e_1, \dots, x_l, e_l\}$. The corresponding labelled rules

$$\frac{\mathsf{x}_1\mathsf{Re}_1, \dots, \mathsf{x}_l\mathsf{Re}_l, \mathsf{w}_1\mathsf{Ro}_1, \dots, \mathsf{w}_m\mathsf{Ro}_m, \Gamma \Rightarrow \Delta}{\mathsf{w}_1\mathsf{Ro}_1, \dots, \mathsf{w}_m\mathsf{Ro}_m, \Gamma \Rightarrow \Delta} \tag{19}$$

with eigenvariables $\mathsf{y}_1, \dots, \mathsf{y}_k$ can, by the subterm property, be restricted as follows: each $\mathsf{x}_j$ and $\mathsf{e}_j$ that is not an eigenvariable must occur in the conclusion sequent.

**Definition 33** (Conmap and premap)**.** Let $\mathcal{M}$ be a Kripke model satisfying (18). An interpretation $[\![\cdot]\!]$ into $\mathcal{M}$ is called a *conmap* for rule (19) if it is a good $\mathcal{M}$-map on the conclusion of (19). A *premap* for the conmap $[\![\cdot]\!]$ is an interpretation $[\![\cdot]\!]^{\vec{y}}_{\vec{\mathsf{y}}}$ that is a good $\mathcal{M}$-map on the premise of (19). Each conmap for (19) has a premap by (18).

**Definition 34** (Interpolable rule)**.** Rule (19) is called *interpolable* for the class $\mathcal{C}_\mathsf{L}$ if its eigenvariables can be ordered $\mathsf{y}_1, \dots, \mathsf{y}_k$ without repetitions in such a way that the following properties hold:

- for each $\mathsf{y}_j$ the antecedent of the premise of (19) contains either $\mathsf{xRy}_j$ for some $\mathsf{x}$ that is not an eigenvariable or $\mathsf{y}_{j'}\mathsf{Ry}_j$ for some $j' < j$. One such "parent," denoted $\mathsf{par}(\mathsf{y}_j)$ is chosen and fixed for each eigenvariable; (*connectedness*) for any $\mathcal{M} = (W, R, V) \in \mathcal{C}_\mathsf{L}$, given any conmap $[\![\cdot]\!]$ into $\mathcal{M}$ for (19) and any premap $[\![\cdot]\!]^{y_1,\dots,y_k}_{\mathsf{y}_1,\dots,\mathsf{y}_k}$ for this conmap, for each $j = 1, \dots, k$

- for any world $y'_j$ such that $y_j R y'_j$, there exists another premap $[\![\cdot]\!]^{y_1,\dots,y_{j-1},y'_j,\dots,y'_k}_{\mathsf{y}_1,\dots,\mathsf{y}_{j-1},\mathsf{y}_j,\dots,\mathsf{y}_k}$ for the conmap $[\![\cdot]\!]$; (*pushability*)
- for arbitrary worlds $z_1, \dots, z_s$ such that $[\![\mathsf{par}(\mathsf{y}_j)]\!]^{y_1,\dots,y_k}_{\mathsf{y}_1,\dots,\mathsf{y}_k} R z_l$ for all $l = 1, \dots, s$, there exists a premap $[\![\cdot]\!]^{y_1,\dots,y_{j-1},y'_j,\dots,y'_k}_{\mathsf{y}_1,\dots,\mathsf{y}_{j-1},\mathsf{y}_j,\dots,\mathsf{y}_k}$ for the conmap $[\![\cdot]\!]$ such that $z_l R y'_j$ for all $l = 1, \dots, s$. (*conjoinability*)

When we say *interpolable rule with eigenvariables* $\mathsf{y}_1, \dots, \mathsf{y}_k$, we imply that connectedness, pushability, and conjoinability are fulfilled for the given order of eigenvariables.

**Definition 35** (Geach properties)**.** The Scott–Lemmon generalizations of the Geach axiom for convergence are known to correspond to the *hijk-convergence properties* $wR^h v \wedge wR^j u \to \exists y(vR^i y \wedge uR^k y)$ [8, Sect. 9]. We avoid degenerate cases by requiring that all $h, i, j, k \geq 1$. In particular, each such $hijk$-convergence property can be easily rewritten as a canonical geometric implication with $i + k - 1$ eigenvariables:

$$\begin{aligned} wRv_1 \wedge \dots \wedge v_{h-1}Rv \quad \wedge \quad wRu_1 \wedge \dots \wedge u_{j-1}Ru \quad \to \\ \exists z_1 \dots \exists z_{i-1} \exists y_1 \dots \exists y_{k-1} \exists y \Big(vRz_1 \wedge \dots \wedge z_{i-1}Ry \quad \wedge \\ uRy_1 \wedge \dots \wedge y_{k-1}Ry\Big)\ . \end{aligned} \tag{20}$$

$$\frac{\mathsf{x}R\mathsf{y}_1, \mathsf{y}_1R\mathsf{y}_2, \dots, \mathsf{y}_{k-1}R\mathsf{y}_k, \mathsf{w}_1R\mathsf{o}_1, \dots, \mathsf{w}_mR\mathsf{o}_m, \Gamma \xRightarrow{\bigovee_{i=1}^{n}\left(\bigowedge_{b=1}^{m_i} \mathsf{u}_{ib}:D_{ib} \owedge \bigowedge_{j=1}^{k} \mathsf{y}_j:C_{ij}\right)} \Delta}{\mathsf{w}_1R\mathsf{o}_1, \dots, \mathsf{w}_mR\mathsf{o}_m, \Gamma \xRightarrow{\bigovee_{i=1}^{n}\left(\bigowedge_{b=1}^{m_i} \mathsf{u}_{ib}:D_{ib} \owedge \mathsf{x}:\Diamond\left(C_{i,1}\wedge\Diamond\left(C_{i,2}\wedge\Diamond\left(\dots\wedge\Diamond(C_{i,k-1}\wedge\Diamond C_{ik})\dots\right)\right)\right)\right)} \Delta}$$

**Table 10.** Interpolating monotelescopic rules. In each rule, $\mathsf{x}$ occurs in the conclusion sequent, none of $\mathsf{y}_j$'s occurs in the conclusion sequent, all $\mathsf{y}_j$ are pairwise distinct, and each $\mathsf{y}_j$ is disctinct from all $\mathsf{u}_{ib}$'s.

It is a bit tedious but not really hard to show the following

**Lemma 36.** *If all frames in $\mathcal{C}_\mathsf{L}$ are $hijk$-convergent and*

- *either transitive or*
- *shift-transitive and $h, j \geq 2$,*

*then the labelled rule generated by the $hijk$-convergence property* (20) *is interpolable for $\mathcal{C}_\mathsf{L}$.*

*Proof.* See Appendix. □

**Lemma 37.** *If for $m < n$ all frames in $\mathcal{C}_\mathsf{L}$ are transitive, Euclidean, and $(n,m)$-transitive ($wR^m x \to wR^n x$), then the labelled sequent rule generated by the $(n,m)$-transitivity property*

$$wRv_1 \wedge \dots \wedge v_{m-1}Rx \to \exists y_1 \dots \exists y_{n-1}(wRy_1 \wedge \dots \wedge y_{n-1}Rx)$$

*is interpolable for $\mathcal{C}_\mathsf{L}$ for the eigenvariable order $\mathsf{y}_1, \dots, \mathsf{y}_{n-1}$.*

**Definition 38** (Interpolant transformation for interpolable rules)**.** For any multiformula in CNF and arbitrary labels $\mathsf{y} \neq \mathsf{x}$, define the following transformation:

$$\mathsf{rem}\left(\mathsf{y}, \mathsf{x}, \bigowedge_{r=1}^{s} \bigovee_{b=1}^{t_r} \mathsf{v}_{rb}:D_{rb}\right) :=$$

$$\bigowedge_{r=1}^{s}\left(\mathsf{x}:\Diamond\Box \bigvee_{\mathsf{v}_{rb}=\mathsf{y}} D_{rb} \quad \ovee \quad \bigovee_{\mathsf{v}_{rb}\neq\mathsf{y}} \mathsf{v}_{rb}:D_{rb}\right) \quad .$$

It is clear that $\mathsf{y}$ does not occur in $\mathsf{rem}(\mathsf{y},\mathsf{x},\mho)$ for any multiformula $\mho$ in CNF. Let a rule R of type (19) be interpolable for $\mathcal{C}_\mathsf{L}$ with eigenvariables $\vec{\mathsf{y}} = \mathsf{y}_1, \dots, \mathsf{y}_k$ and $\mho$ be a multiformula in CNF. For each $j = 0, \dots, k$, we recursively define

$$\mathsf{rem}_j(\vec{\mathsf{y}}, R, \mho) := \begin{cases} \mho & \text{if } j = k, \\ \mathsf{rem}\Big(\mathsf{y}_{j+1}, \mathsf{par}(\mathsf{y}_{j+1}), \mathsf{rem}_{j+1}(\vec{\mathsf{y}}, R, \mho)\Big) & \text{if } j \leq k-1. \end{cases} \tag{21}$$

Note that $\mathsf{rem}_j(\vec{\mathsf{y}}, R, \mho)$ is in CNF and does not contain occurrences of $\mathsf{y}_{j+1}, \dots, \mathsf{y}_k$ by connectedness of $R$. Finally, we define

$$\mathsf{rem}(\vec{\mathsf{y}}, R, \mho) := \mathsf{rem}_0(\vec{\mathsf{y}}, R, \mho) \quad ,$$

which does not contain any eigenvariables of $R$.

Our goal is to show that for a rule $R$ of type (19),

$$\frac{\mathsf{x}_1R\mathsf{e}_1, \dots, \mathsf{x}_lR\mathsf{e}_l, \mathsf{w}_1R\mathsf{o}_1, \dots, \mathsf{w}_mR\mathsf{o}_m, \Gamma \overset{\mho}{\Rightarrow} \Delta}{\mathsf{w}_1R\mathsf{o}_1, \dots, \mathsf{w}_mR\mathsf{o}_m, \Gamma \xRightarrow{\mathsf{rem}(\vec{\mathsf{y}},R,\mho)} \Delta}\, R \tag{22}$$

for any multiformula $\mho$ in CNF.

**Lemma 39.** *Let rule $R$ of type* (19) *be interpolable for $\mathcal{C}_\mathsf{L}$ with eigenvariables $\mathsf{y}_1, \dots, \mathsf{y}_k$ and $\mho$ be a multiformula in CNF that satisfies* (1) *for the premise of $R$. Let all models from $\mathcal{C}_\mathsf{L}$ satisfy the corresponding frame condition* (18) *and $\mathcal{M} \in \mathcal{C}_\mathsf{L}$. Let $[\![\cdot]\!]$ be a conmap into $\mathcal{M}$ for R. Then*

$$\mathcal{M} \vDash [\![\mathsf{Ant}(\Gamma)]\!] \qquad \textit{implies} \qquad \mathcal{M} \vDash [\![\mathsf{rem}(\vec{\mathsf{y}}, R, \mho)]\!] \quad .$$

*Proof.* Let us use the abbreviation $\mho_j := \mathsf{rem}_j(\vec{\mathsf{y}}, R, \mho)$. Assume that $\mathcal{M} \vDash [\![\mathsf{Ant}(\Gamma)]\!]$. We prove by induction on $j = k, \dots, 0$ that $\mathcal{M} \vDash [\![\mho_j]\!]^{y_1,\dots,y_j}_{\mathsf{y}_1,\dots,\mathsf{y}_j}$ for any premap $[\![\cdot]\!]^{\vec{y}}_{\vec{\mathsf{y}}}$ for the conmap $[\![\cdot]\!]$. The base case $j = k$ follows from (1) because $\mho_k = \mho$ and because $\mathsf{Ant}(\cdot)$ is insensitive to relational atoms.

Assume the IH holds for $j = i$. Consider an arbitrary premap $[\![\cdot]\!]^{\vec{y}}_{\vec{\mathsf{y}}}$ for $[\![\cdot]\!]$. By pushability of $R$, for any $y'_i$ such that $y_iRy'_i$, there is another premap $[\![\cdot]\!]^{y_1,\dots,y_{i-1},y'_i,\dots,y'_k}_{\mathsf{y}_1,\dots,\mathsf{y}_{i-1},\mathsf{y}_i,\dots\mathsf{y}_k}$ for $[\![\cdot]\!]$. Thus, by IH, for each conjunct $\bigovee_{b=1}^{t} \mathsf{v}_b : D_b$ of $\mho_i$,

$$y_iRy'_i \qquad \text{implies} \qquad \mathcal{M} \vDash \left[\!\!\left[\bigovee_{b=1}^{t} \mathsf{v}_b : D_b\right]\!\!\right]^{y_1,\dots,y_{i-1},y'_i}_{\mathsf{y}_1,\dots,\mathsf{y}_{i-1},\mathsf{y}_i} \quad .$$

We need to show that

$$\mathcal{M} \vDash \left[\!\!\left[\mathsf{par}(\mathsf{y}_i) : \Diamond\Box \bigvee_{\mathsf{v}_b=\mathsf{y}_i} D_b \ovee \bigovee_{\mathsf{v}_b\neq\mathsf{y}_i} \mathsf{v}_b : D_b\right]\!\!\right]^{y_1,\dots,y_{i-1}}_{\mathsf{y}_1,\dots,\mathsf{y}_{i-1}} \tag{23}$$

for the corresponding conjunct of $\mho_{i-1}$. We only show this for conjuncts with $\mathcal{M}, [\![\mathsf{v}_b]\!]^{y_1,\dots,y_{i-1}}_{\mathsf{y}_1,\dots,\mathsf{y}_{i-1}} \nVdash D_b$ for all $\mathsf{v}_b \neq \mathsf{y}_i$ (the case for the remaining conjuncts is trivial). For these conjuncts,

$$y_iRy'_i \qquad \text{implies} \qquad \mathcal{M}, y'_i \Vdash \bigvee_{\mathsf{v}_b=\mathsf{y}_i} D_b \ ,$$

i.e., $\mathcal{M}, y_i \Vdash \Box \bigvee_{\mathsf{v}_b=\mathsf{y}_i} D_b$. Given that $[\![\cdot]\!]^{y_1,\dots,y_{i-1},y_i}_{\mathsf{y}_1,\dots,\mathsf{y}_{i-1},\mathsf{y}_i}$ agrees with the premap $[\![\cdot]\!]^{\vec{y}}_{\vec{\mathsf{y}}}$ on both $\mathsf{y}_i$ and $\mathsf{par}(\mathsf{y}_i)$ and that $[\![\cdot]\!]^{y_1,\dots,y_{i-1}}_{\mathsf{y}_1,\dots,\mathsf{y}_{i-1}}$ agrees with them on $\mathsf{par}(\mathsf{y}_i)$, it follows by connectedness that $[\![\mathsf{par}(\mathsf{y}_i)]\!]^{y_1,\dots,y_{i-1}}_{\mathsf{y}_1,\dots,\mathsf{y}_{i-1}} Ry_i$. Thus, for these conjuncts, (23) is true because $\mathcal{M}, [\![\mathsf{par}(\mathsf{y}_i)]\!]^{y_1,\dots,y_{i-1}}_{\mathsf{y}_1,\dots,\mathsf{y}_{i-1}} \Vdash \Diamond\Box \bigvee_{\mathsf{v}_b=\mathsf{y}_i} D_b$. Since all conjuncts of $\mho_{i-1}$ are forced by $[\![\cdot]\!]^{y_1,\dots,y_{i-1}}_{\mathsf{y}_1,\dots,\mathsf{y}_{i-1}}$, it follows that $\mho_{i-1}$ is forced. This completes the induction proof.

In particular, $\mathcal{M} \vDash [\![\mho_0]\!]$ for any premap $[\![\cdot]\!]^{\vec{y}}_{\vec{\mathsf{y}}}$. It remains to note that such premaps exist by (18) and that $\mathsf{rem}(\vec{\mathsf{y}}, R, \mho) = \mho_0$. □

**Lemma 40.** *Let rule $R$ of type* (19) *be interpolable for $\mathcal{C}_\mathsf{L}$ with eigenvariables $\mathsf{y}_1, \dots, \mathsf{y}_k$. Let $\mho$ be a multiformula in CNF. Let all models from $\mathcal{C}_\mathsf{L}$ satisfy the corresponding frame condition* (18) *and $\mathcal{M} \in \mathcal{C}_\mathsf{L}$. Let $[\![\cdot]\!]$ be a conmap into $\mathcal{M}$ for R. Then*

$$\mathcal{M} \vDash [\![\mathsf{rem}(\vec{\mathsf{y}}, R, \mho)]\!] \qquad \textit{implies} \qquad \mathcal{M} \vDash [\![\mho]\!]^{\vec{y}}_{\vec{\mathsf{y}}}$$

*for some premap $[\![\cdot]\!]^{\vec{y}}_{\vec{\mathsf{y}}}$ for $[\![\cdot]\!]$.*

*Proof.* Once again, let $\mho_j := \mathsf{rem}_j(\vec{\mathsf{y}}, R, \mho)$. We prove by induction on $j = 0, \dots, k$ that there exists a premap $[\![\cdot]\!]^{y^j_1,\dots,y^j_j,y^j_{j+1},\dots,y^j_k}_{\mathsf{y}_1,\dots,\mathsf{y}_j,\mathsf{y}_{j+1},\dots,\mathsf{y}_k}$

for $[\![\cdot]\!]$ such that $\mathcal{M} \vDash [\![\mho_j]\!]^{y_1^1,\dots,y_j^j}_{\mathsf{y}_1,\dots,\mathsf{y}_j}$. The base case $j = 0$ follows from (18) because $\mathsf{rem}(\vec{\mathsf{y}}, R, \mho) = \mho_0$. Assume by the IH that $\mathcal{M} \vDash [\![\mho_i]\!]^{y_1^1,\dots,y_i^i}_{\mathsf{y}_1,\dots,\mathsf{y}_i}$ for a premap $[\![\cdot]\!]^{y_1^1,\dots,y_i^i,y_{i+1}^i,\dots,y_k^i}_{\mathsf{y}_1,\dots,\mathsf{y}_i,\mathsf{y}_{i+1},\dots,\mathsf{y}_k}$. In particular,

$$\mathcal{M} \vDash \left[\!\!\left[\mathsf{par}(\mathsf{y}_{i+1}) : \Diamond\Box \bigvee_{\mathsf{v}_{rb}=\mathsf{y}_{i+1}} D_{rb} \oslash \bigoslash_{\mathsf{v}_{rb}\neq\mathsf{y}_{i+1}} \mathsf{v}_{rb} : D_{rb}\right]\!\!\right]^{y_1^1,\dots,y_i^i}_{\mathsf{y}_1,\dots,\mathsf{y}_i}$$

for the $r$th conjunct $\bigoslash_{b=1}^{t_r} \mathsf{v}_{rb} : D_{rb}$ of $\mho_{i+1}$. If for some $b = 1, \dots, t_r$ with $\mathsf{v}_{rb} \neq \mathsf{y}_{i+1}$, we have $\mathcal{M}, [\![\mathsf{v}_{rb}]\!]^{y_1^1,\dots,y_i^i}_{\mathsf{y}_1,\dots,\mathsf{y}_i} \Vdash D_{rb}$, then

$$\mathcal{M} \vDash \left[\!\!\left[\bigoslash_{b=1}^{t_r} \mathsf{v}_{rb} : D_{rb}\right]\!\!\right]^{y_1^1,\dots,y_i^i,z}_{\mathsf{y}_1,\dots,\mathsf{y}_i,\mathsf{y}_{i+1}} \qquad \text{for any world } z \ . \quad (24)$$

Let $r_1, \dots, r_s$ be the numbers of the remaining conjuncts for which for all $l = 1, \dots, s$

$$\mathcal{M}, [\![\mathsf{par}(\mathsf{y}_{i+1})]\!]^{y_1^1,\dots,y_i^i}_{\mathsf{y}_1,\dots,\mathsf{y}_i} \Vdash \Diamond\Box \bigvee_{\mathsf{v}_{r_l,b}=\mathsf{y}_{i+1}} D_{r_l,b} \ .$$

If $s = 0$, then all conjuncts of $\mho_{i+1}$ are covered by (24), so we can keep the same premap and set $y_{i+1}^{i+1} := y_{i+1}^i$. Otherwise, there exist worlds $z_1, \dots, z_s$ such that for all $l = 1, \dots, s$ we have $[\![\mathsf{par}(\mathsf{y}_{i+1})]\!]^{y_1^1,\dots,y_i^i}_{\mathsf{y}_1,\dots,\mathsf{y}_i} R z_l$ and $\mathcal{M}, z_l \Vdash \Box \bigvee_{\mathsf{v}_{r_l,b}=\mathsf{y}_{i+1}} D_{r_l,b}$. By connectedness, the same holds for the premap $[\![\cdot]\!]^{y_1^1,\dots,y_i^i,y_{i+1}^i,\dots,y_k^i}_{\mathsf{y}_1,\dots,\mathsf{y}_i,\mathsf{y}_{i+1},\dots,\mathsf{y}_k}$. By conjoinability, there exists a premap $[\![\cdot]\!]^{y_1^1,\dots,y_i^i,y_{i+1}^{i+1},y_{i+2}^{i+1},\dots,y_k^{i+1}}_{\mathsf{y}_1,\dots,\mathsf{y}_i,\mathsf{y}_{i+1},\mathsf{y}_{i+2},\dots,\mathsf{y}_k}$ such that $z_l R y_{i+1}^{i+1}$ for all $l = 1, \dots, s$. Moreover, for this $y_{i+1}^{i+1}$ we have $\mathcal{M}, y_{i+1}^{i+1} \Vdash \bigvee_{\mathsf{v}_{r_l,b}=\mathsf{y}_{i+1}} D_{r_l,b}$ for all $l = 1, \dots, s$, i.e.,

$$\mathcal{M}, [\![\mathsf{v}_{r_l,b_l}]\!]^{y_1^1,\dots,y_i^i,y_{i+1}^{i+1}}_{\mathsf{y}_1,\dots,\mathsf{y}_i,\mathsf{y}_{i+1}} \Vdash D_{r_l,b_l} \quad (25)$$

for some $\mathsf{v}_{r_l,b_l} = \mathsf{y}_{i+1}$. Thus, for $[\![\cdot]\!]^{y_1^1,\dots,y_i^i,y_{i+1}^{i+1}}_{\mathsf{y}_1,\dots,\mathsf{y}_i,\mathsf{y}_{i+1}}$, these $s$ conjuncts of $\mho_{i+1}$ are forced due to (25) and the remaining conjuncts are forced by (24) for $z = y_{i+1}^{i+1}$. This completes the induction proof.

Since $\mho = \mho_k$, the desired statement holds for $\vec{y} = y_1^1, \dots, y_k^k$. □

**Theorem 41.** *Let rule $R$ of type* (19) *satisfy the subterm property and be interpolable for $\mathcal{C}_\mathsf{L}$ with eigenvariables $\mathsf{y}_1, \dots, \mathsf{y}_k$ and $\mho$ be an interpolant for the premise of $R$. Let all models from $\mathcal{C}_\mathsf{L}$ satisfy the corresponding frame condition* (18)*. Then* (22) *is the interpolant transformation for $R$.*

*Proof.* We know that no eigenvariables of $R$ occur in $\mathsf{rem}(\vec{\mathsf{y}}, R, \mho)$. All the remaining labels from it must occur in the conclusion of $R$ by the subterm property. Since $R$ does not change the common language (no labelled formula is changed) and since $\mathsf{rem}(\vec{\mathsf{y}}, R, \mho)$ has the same propositional atoms (with the same polarities) as $\mho$, the common language condition is also fulfilled. Consider an arbitrary model $\mathcal{M} \in \mathcal{C}_\mathsf{L}$ and an arbitrary conmap for $R$. Since $\mho$ is an interpolant for the premise, it satisfies (1) for it. Assume that $\mathcal{M} \vDash [\![\mathsf{Ant}(\mathsf{w}_1 \mathsf{R} \mathsf{o}_1, \dots, \mathsf{w}_m \mathsf{R} \mathsf{o}_m, \Gamma)]\!]$, which is the same as $\mathcal{M} \vDash [\![\mathsf{Ant}(\Gamma)]\!]$. By Lemma 39, $\mathcal{M} \vDash [\![\mathsf{rem}(\vec{\mathsf{y}}, R, \mho)]\!]$. This completes the proof of (1) for the conclusion of $R$. Assume now that $\mathcal{M} \vDash [\![\mathsf{rem}(\vec{\mathsf{y}}, R, \mho)]\!]$. By Lemma 40, there is a premap $[\![\cdot]\!]^{\vec{y}}_{\vec{\mathsf{y}}}$ for $[\![\cdot]\!]$ such that $\mathcal{M} \vDash [\![\mho]\!]^{\vec{y}}_{\vec{\mathsf{y}}}$. Since $\mho$ is an interpolant for the premise, it follows that $\mathcal{M} \vDash [\![\mathsf{Cons}(\Delta)]\!]^{\vec{y}}_{\vec{\mathsf{y}}}$. But $\Delta$ contains no eigenvariables. Hence, $\mathcal{M} \vDash [\![\mathsf{Cons}(\Delta)]\!]$. This completes the proof of (2) and that $\mho_0$ is an interpolant for the conclusion of $R$. □

In the following corollary we collect all the interpolation results obtainable by the methods of this paper:

**Corollary 42.** *Modal logics enjoy both CIP and LIP if they are complete w.r.t. the class of Kripke models defined by any combination of the following properties*

- *reflexivity, transitivity, symmetry, seriality, and Euclideanness;*
- *shift reflexivity, shift transitivity, shift symmetry, shift seriality, and shift Euclideanness;*
- *any property obtained by replacing the "shift" condition above by an arbitrary conjunction of relational atoms;*
- *functionality;*
- *$(1, n)$-transitivity;*
- *strictly irreflexive, strictly reflexive, or unspecified discreteness of the frame;*
- *$hijk$-convergence with $h, i, j, k \geq 1$ (in presence of transitivity);*
- *$hijk$-convergence with $i, k \geq 1$ and $h, j \geq 2$ (in presence of shift transitivity);*
- *density (in presence of transitivity and Euclideanness);*
- *$(n, m)$-transitivity for $m < n$ (in presence of transitivity and Euclideanness).*[7]

*In particular, the list of logics with CIP/LIP proved using labelled sequents includes all 15 logics of the so-called modal cube from [8, Sect. 8],* K4.2*,* S4.2*,* Triv*,* Verum*, and* $\mathsf{K4}1_{1,n}$ *from [3, Table 4.2], as well as the infinite family of non-degenerate Geach logics over* K4 *and almost the full family of Geach logics over* K5 *(due to the shift transitivity of the latter).*

*Remark* 43. While cyclical relationships among the eigenvariables are not forbidden, they certainly make fulfilling the pushability and conjoinability criteria problematic. Another problem is the presence of *uplinks* $\mathsf{y}R\mathsf{w}$ from an eigenvariable $\mathsf{y}$ to a label $\mathsf{w}$ from the conclusion of the rule. For instance, density can, at the moment, only be processed in presence of rather strong additional properties.

## 7. Related Work

The body of work on interpolation is so great and so varied that it is hopeless to try giving even a restricted overview of the field. To the best of our knowledge, this is the first result on proving interpolation using labelled sequent calculi. One should, however, acknowledge several recent advances in other proof formalisms. Bílková [1] uses nested sequent calculi to show the stronger uniform interpolation for several modal logics. Pattinson [15] provides a blanket proof of uniform interpolation for the class of rank-1 modal logics, spanning multiple types of modal logics, but somewhat restricted as far as normal modal logics are concerned: for instance, it does not include any transitive logics. Iemhoff [9] connects the existence of ordinary sequent calculi of particular type to the property of uniform interpolation, which can be used to show the absence of such sequent calculi, but can only prove uniform interpolation for logics with sequent systems. Brotherston and Goré [2] pursue an endeavor similar to the one undertaken in this paper for display calculi, which are closer related to algebraic semantics and require working in an extended language causing potential problems in showing conservativity and decidability.

[7] $(n, m)$-transitivity for $m \geq n$ follows from transitivity.

## 8. Conclusion and Future Work

We have developed a novel constructive and modular method of proving the Lyndon (and by extention Craig) Interpolation Property for modal logics by using symmetric labelled sequent calculi. The method is sufficient to establish the LIP for all frame conditions described by quantifier-free Horn formulas. For geometric formulas, the method generally requires additional conditions similar to transitivity and convergence in nature, but is still sufficient to tackle many of the standard modal logics.

There are still many questions to be answered. There exist general theorems establishing interpolation semantically but not constructively. It would be interesting to compare their semantic restrictions with those of our method. The extension to multimodal logics should be unproblematic. Also the extension to first-order languages is long overdue. The use of symmetric sequents in this paper masks the difficulties of approaching intuitionistically-grounded theories. Logics like $\mathsf{GL}$ can be captured by labelled sequents even though they are not first-order definable. It stands to reason that our method extends to them too.

## Acknowledgments

I am grateful to Melvin Fitting, whose idea started the whole interpolation project. I thank Sara Negri for encouraging me to tackle labelled sequents (and apologize for not doing it earlier). I am deeply in debt to Björn Lellmann, who is always ready to listen to my ramblings and has provided innumerable inspiring suggestions (and readily available knowledge) for improving this paper. I am also privileged to have such wonderful colleagues who make my work great fun.

## A. Appendix

In this appendix, we prove certain facts that belong more to graph theory than to a study of interpolation. We start with auxiliary lemmas that would shorten the graph-theoretic arguments.

**Proposition 1.** *Let $\mathcal{F} = (W, R)$ be a Kripke frame.*
*If $R$ is transitive, then $R^k \subseteq R^l$ whenever $k > l \geq 1$.*
*If $R$ is shift-transitive, then $R^k \subseteq R^l$ whenever $k > l \geq 2$.*
*If $R$ is shift-transitive and $k > l \geq 1$, then $xRy$ and $yR^k z$ imply $yR^l z$.*

*Proof.* Trivial. □

**Proposition 2.** *Let a Kripke frame $\mathcal{F} = (W, R)$ enjoy the $hjik$-convergence property and either shift transitivity if $h, j \geq 2$ or transitivity otherwise. Then $\mathcal{F}$ enjoys the following Geach-shift convergence properties: for any $a \geq 0$,*

$$wR^{h+a}v \wedge wR^j u \wedge \bigwedge_{s=1}^{S} vRx_s \to \exists y \left(\bigwedge_{s=1}^{S} x_s Ry\right) \quad ;$$

$$wR^h v \wedge wR^{j+a} u \wedge \bigwedge_{s=1}^{S} uRx_s \to \exists y \left(\bigwedge_{s=1}^{S} x_s Ry\right) \quad .$$

*Proof.* We only prove the first property; the proof of the second is analogous. For each $1 \leq s \leq S$, we have $wR^{h+a}vRx_s$. Thus, $wR^h x_s$ by Prop. 1. Define $y_0 := u$. By induction on $s = 1, \ldots, S$ we find worlds $y_s$ such that $x_s R^i y_s$ and $wR^j y_s$ as follows. Note that $wR^j y_0$. By $hjik$-convergence for $wR^h x_s$ and $wR^j y_{s-1}$, there exists $y_s$ such that $x_s R^i y_s$ and $y_{s-1} R^k y_s$. From $wR^j y_{s-1} R^k y_s$ we get $wR^j y_s$ by Prop. 1. For the last world $y_S$ and any $1 \leq s \leq S$ we have $x_s R^i y_s R^k y_{s+1} R^k \ldots R^k y_S$. Thus, by shift transitivity, $x_s R y_S$ for all $1 \leq s \leq S$. We set $y := y_S$. □

***Proof of Lemma 36.*** The following abbreviations are useful: $v_0 = u_0 := w$, $v_h = z_0 := v$, $u_j = y_0 := u$, and $z_i = y_k := y$.

**Connectedness** holds for the order

$$\langle \mathsf{z}_1, \ldots, \mathsf{z}_{i-1}, \mathsf{y}_1, \ldots, \mathsf{y}_{k-1}, \mathsf{y} \rangle$$

on eigenvariables given by the existential quantifiers in (20) if we define $\mathsf{par}(\mathsf{y}_s) := \mathsf{y}_{s-1}$ for $1 \leq s \leq k$ and $\mathsf{par}(\mathsf{z}_s) := \mathsf{z}_{s-1}$ for $1 \leq s \leq i-1$ (in particular, $\mathsf{par}(\mathsf{y}) = \mathsf{par}(\mathsf{y}_k) = \mathsf{y}_{k-1}$, $\mathsf{par}(\mathsf{y}_1) = \mathsf{y}_0 = \mathsf{u}$, and $\mathsf{par}(\mathsf{z}_1) = \mathsf{z}_0 = \mathsf{v}$).

**Pushability.** We are given a conmap $[\![\cdot]\!]$ for the Geach rule. In particular, $[\![\mathsf{w}_s]\!] R [\![\mathsf{w}_{s+1}]\!]$ for all $0 \leq s \leq h-1$ and $[\![\mathsf{u}_s]\!] R [\![\mathsf{u}_{s+1}]\!]$ for all $0 \leq s \leq j-1$. Since it does not matter what $[\![\cdot]\!]$ does on eigenvariables, for we assume w.l.o.g. that $[\![\cdot]\!]$ is also a premap, i.e., that, in addition, $[\![\mathsf{z}_s]\!] R [\![\mathsf{z}_{s+1}]\!]$ for all $0 \leq s \leq i-1$ and $[\![\mathsf{y}_s]\!] R [\![\mathsf{y}_{s+1}]\!]$ for all $0 \leq s \leq k-1$.

Let $[\![y]\!] R y'$. We need to show that making $[\![y]\!]' = y'$ without changing interpretations of any other worlds yields another premap. This follows immediately from (shift) transitivity: $[\![\mathsf{z}_{i-1}]\!] R [\![\mathsf{y}]\!] R y'$ implies $[\![\mathsf{z}_{i-1}]\!] R y'$ and $[\![\mathsf{y}_{k-1}]\!] R [\![\mathsf{y}]\!] R y'$ implies $[\![\mathsf{y}_{k-1}]\!] R y'$.

Let $[\![\mathsf{y}_s]\!] R y'_s$ for some $1 \leq s \leq k-1$. We need to construct an alternative premap $[\![\cdot]\!]'$ with $[\![\mathsf{y}_s]\!]' = y'_s$ by changing only the worlds interpreting $\mathsf{y}_{s+1}, \ldots, \mathsf{y}_k = \mathsf{y}$. Note that $[\![\mathsf{y}_{s-1}]\!] R [\![\mathsf{y}_s]\!] R y'_s$, making $[\![\mathsf{y}_{s-1}]\!] R y'_s$ by (shift) transitivity. Since $[\![\mathsf{w}]\!] R^h [\![\mathsf{v}]\!] R^{i-1} [\![\mathsf{z}_{i-1}]\!]$, it follows by Prop. 1 that $[\![\mathsf{w}]\!] R^h [\![\mathsf{z}_{i-1}]\!]$ (note that shift transitivity is not sufficient here if $h = 1$). Since $[\![\mathsf{w}]\!] R^j [\![\mathsf{u}]\!] R^s [\![\mathsf{y}_s]\!] R y'_s$, it follows by Prop. 1 that $[\![\mathsf{w}]\!] R^j y'_s$ (note that shift transitivity is not sufficient here if $j = 1$). By (20), there exists $y'$ such that $[\![\mathsf{z}_{i-1}]\!] R^i y'$ and $y'_s R^k y'$. Using Prop. 1, we conclude $[\![\mathsf{z}_{i-1}]\!] R y'$ from the former and $y'_s R^{k-s} y'$ from the latter (note that $k - s > 0$). Thus, there exist worlds such that $y'_s R y'_{s+1} R \ldots y'_{k-1} R y'$, and we put $[\![\mathsf{y}_{s+a}]\!]' := y'_{s+a}$ for $1 \leq a \leq k-s-1$ and $[\![\mathsf{y}]\!]' := y'$.

Let $[\![\mathsf{z}_s]\!] R z'_s$ for some $1 \leq s \leq i-1$. We need to construct an alternative premap $[\![\cdot]\!]'$ with $[\![\mathsf{z}_s]\!]' = z'_s$ by changing only the worlds interpreting $\mathsf{z}_{s+1}, \ldots, \mathsf{z}_{i-1}, \mathsf{y}_1, \ldots, \mathsf{y}_k = \mathsf{y}$. Note that $[\![\mathsf{z}_{s-1}]\!] R [\![\mathsf{z}_s]\!] R z'_s$, making $[\![\mathsf{z}_{s-1}]\!] R z'_s$ by (shift) transitivity. Since $[\![\mathsf{w}]\!] R^h [\![\mathsf{v}]\!] R^s [\![\mathsf{z}_s]\!] R z'_s$, it follows by Prop. 1 that $[\![\mathsf{w}]\!] R^h z'_s$. By (20) from this and $[\![\mathsf{w}]\!] R^j [\![\mathsf{u}]\!]$, there exists $y'$ such that $z'_s R^i y'$ and $[\![\mathsf{u}]\!] R^k y'$. Using Prop. 1, we conclude $z'_s R^{i-s} y'$ from the former (note that $i - s > 0$). Thus, there exist worlds such that $z'_s R z'_{s+1} R \ldots z'_{i-1} R y'$ and $[\![\mathsf{u}]\!] R y'_1 R \ldots R y'_{k-1} R y'$, and we put $[\![\mathsf{z}_{s+a}]\!]' := z'_{s+a}$ for $1 \leq a \leq i-s-1$, $[\![\mathsf{y}_a]\!]' := y'_a$ for $1 \leq a \leq k-1$, and $[\![\mathsf{y}]\!]' := y'$.

**Conjoinability.** We are working with the same conmap/premap $[\![\cdot]\!]$ for the Geach rule.

Let for some $1 \leq s \leq i-1$, we have $[\![\mathsf{z}_{s-1}]\!] R x_t$ for all $1 \leq t \leq T$. Without changing interpretations of $\mathsf{z}_1, \ldots, \mathsf{z}_{s-1}$, we need to find another premap with $[\![\mathsf{z}_s]\!]' = z'_s$ such that $x_t R z'_s$ for all $1 \leq t \leq T$. From $[\![\mathsf{w}]\!] R^{h+s-1} [\![\mathsf{z}_{s-1}]\!]$, $[\![\mathsf{w}]\!] R^j [\![\mathsf{u}]\!]$, $[\![\mathsf{z}_{s-1}]\!] R x_t$ for $1 \leq t \leq T$, and $[\![\mathsf{z}_{s-1}]\!] R [\![\mathsf{z}_s]\!]$, it follows by Prop. 2 that there is $z'_s$ such that $x_t R z'_s$ for all $1 \leq t \leq T$ and $[\![\mathsf{z}_s]\!] R z'_s$. By pushability, it follows that there is a premap mapping $\mathsf{z}_s$ to $z'_s$ and not affecting eigenvariables preceding $\mathsf{z}_s$ in the eigenvariable ordering.

Let for some $1 \leq s \leq k$, we have $[\![\mathsf{y}_{s-1}]\!] R x_t$ for all $1 \leq t \leq T$. Without changing interpretations of $\mathsf{z}_1, \ldots, \mathsf{z}_{i-1}, \mathsf{y}_1, \ldots, \mathsf{y}_{s-1}$, we need to find another premap with $[\![\mathsf{y}_s]\!]' = y'_s$ such that $x_t R y'_s$ for all $1 \leq t \leq T$. From $[\![\mathsf{w}]\!] R^h [\![\mathsf{v}]\!]$, $[\![\mathsf{w}]\!] R^{j+s-1} [\![\mathsf{y}_{s-1}]\!]$, $[\![\mathsf{y}_{s-1}]\!] R x_t$ for $1 \leq t \leq T$, and $[\![\mathsf{y}_{s-1}]\!] R [\![\mathsf{y}_s]\!]$, it follows by Prop. 2 that there is $y'_s$ such that $x_t R y'_s$ for all $1 \leq t \leq T$ and $[\![y_s]\!] R y'_s$. By pushability, it follows that there is a premap mapping $\mathsf{y}_s$ to $y'_s$ and not affecting eigenvariables preceding $\mathsf{y}_s$ in the eigenvariable ordering. □